\documentclass[journal]{IEEEtran}

\usepackage[noadjust]{cite}
\usepackage{multirow}
\usepackage{multicol}
\usepackage{amsmath}
\usepackage{graphicx}
\usepackage[table]{xcolor}
\usepackage[colorlinks=true, allcolors=black]{hyperref}
\usepackage{hyperref}
\usepackage{orcidlink}
\usepackage{booktabs} 
\usepackage{comment}
\usepackage[siunitx]{circuitikz}
\usepackage{caption}
\usepackage{subcaption}
\usepackage{amssymb}
\usepackage{tikz}
\usepackage{siunitx}
\usepackage{textcomp}

\usepackage[acronym]{glossaries}
\newacronym{anova}{ANOVA}{analysis of variance}
\newacronym{artemis}{ARTEMIS}{Assessment and Reliability of Transport Emission Models and Inventory Systems}
\newacronym{cad}{CAD}{computer-aided design}
\newacronym{dgsm}{DGSM}{derivative-based global sensitivity measures}
\newacronym{ecm}{ECM}{equivalent circuit model}
\newacronym{fem}{FEM}{finite element method}
\newacronym{iga}{IGA}{isogeometric analysis}
\newacronym{im}{IM}{induction machine}
\newacronym{kle}{KLE}{Karhunen-L\`ove expansion}
\newacronym{ml}{ML}{machine learning}
\newacronym{mcs}{MCS}{Monte Carlo sampling}
\newacronym{mae}{MAE}{mean absolute error}
\newacronym{mse}{MSE}{mean square error}
\newacronym{pce}{PCE}{polynomial chaos expansion}
\newacronym{pdf}{PDF}{probability density function}
\newacronym{pmsm}{PMSM}{permanent magnet synchronous machine}
\newacronym[longplural={quantities of interest}]{qoi}{QoI}{quantity of interest}
\newacronym{rv}{RV}{random variable}
\newacronym{gsa}{GSA}{global sensitivity analysis}
\newacronym{wltp}{WLTP}{Worldwide Harmonised Light vehicles Test Procedure}
\newacronym{uq}{UQ}{uncertainty quantification}

\begin{document}

\title{Multivariate Sensitivity Analysis of Electric Machine Efficiency Maps and Profiles Under Design Uncertainty}

\author{
\IEEEauthorblockN{
Aylar Partovizadeh\IEEEauthorrefmark{1}\orcidlink{0009-0008-1894-2203},
Sebastian Sch\"ops\IEEEauthorrefmark{1}\orcidlink{0000-0001-9150-0219},
Dimitrios Loukrezis\IEEEauthorrefmark{2}\orcidlink{0000-0003-1264-1182}\\
}
\IEEEauthorblockA{\IEEEauthorrefmark{1}Computational Electromagnetics Group, Technische Universit\"at Darmstadt, 64289 Darmstadt, Germany}
\IEEEauthorblockA{\IEEEauthorrefmark{2}Scientific Computing, Centrum Wiskunde \& Informatica (CWI), 1098~XG Amsterdam, The Netherlands}
\thanks{Corresponding authors: Aylar Partovizadeh (email: aylar.partovizadeh@tu-darmstadt.de); Dimitrios Loukrezis (email: d.loukrezis@cwi.nl).}
}


\maketitle

\begin{abstract}
This work introduces the use of multivariate global sensitivity analysis for assessing the impact of uncertain electric machine design parameters on efficiency maps and profiles. 
Contrary to the common approach of applying variance-based (Sobol') sensitivity analysis elementwise, multivariate sensitivity analysis provides a single sensitivity index per parameter, thus allowing for a holistic estimation of parameter importance over the full efficiency map or profile. 
Its benefits are demonstrated on permanent magnet synchronous machine models of different fidelity. 
Computations based on Monte Carlo sampling and polynomial chaos expansions are compared in terms of computational cost. 
The sensitivity analysis results are subsequently used to simplify the models, by fixing non-influential parameters to their nominal values and allowing random variations only for influential parameters. 
Uncertainty estimates obtained with the full and reduced models confirm the validity of model simplification guided by multivariate sensitivity analysis.
\end{abstract}

\begin{IEEEkeywords}
\textcolor{black}{electric machines}, \textcolor{black}{permanent magnet machines}, \textcolor{black}{sensitivity analysis}, \textcolor{black}{uncertain systems}, \textcolor{black}{vehicle driving}.
\end{IEEEkeywords}

\section{Introduction}
\label{sec:intro}
\IEEEPARstart{T}{he} advancement of electric mobility hinges on improving the efficiency of electric machines \cite{heineke2024spotlight}. 
This goal concentrates significant research and innovation efforts and places efficiency among the most important \glspl{qoi} in modern electric machine design~\cite{gobbi2024traction}.
Efficiency maps describing the maximum efficiency over the full torque-speed operation range, as well as efficiency profiles corresponding to specific driving cycles, are particularly important tools for assessing the performance of an electric machine for varying and dynamic operating conditions~\cite{roshandel2022efficiency, dhakal2025comparative}.

During the design phase, efficiency maps and profiles are commonly estimated with the help of simulation models, ranging from relatively simple \glspl{ecm} to detailed \gls{cad} models solved with advanced numerical techniques.
However, irrespective of resolution capability and approximation accuracy, simulation models are simplified and idealized representations of real-world devices. 
As such, they do not account for modeling inaccuracies, manufacturing tolerances, material imperfections, or other factors that may arise during the design, fabrication or operation of an electric machine.
Taking these uncertainties into consideration and assessing their impact upon crucial \glspl{qoi} like efficiency is a significant design challenge.

To that end, \gls{uq} studies are nowadays an important part of model-based electric machine design processes~\cite{ramarotafika2012stochastic, fratila2013stochastic, clenet2013uncertainty, offermann2015uncertainty, bontinck2015response, belahcen2015uncertainty, liu2015study, galetzka2019multilevel, beltran2020uncertainty, lei2020robust, yang2022computationally,  liu2022permanent, liang2024sensitivity, partovizadeh2025fourier}.
One particularly important study is \gls{gsa}, which allows to quantitatively attribute the uncertainty observed in a \gls{qoi} to the uncertain design parameters~\cite{saltelli2008global, razavi2021future}. 
\gls{gsa} results  can guide design changes and adjustments, for example, by fixing non-influential parameters and concentrating design efforts on the parameters mainly responsible for \gls{qoi} variations.
Sobol' \gls{gsa} is a popular variance-based approach, which quantifies the impact of the uncertain parameters and their combinations upon the \gls{qoi}'s variability in the form of numerical values known as Sobol' indices~\cite{sobol2001global}.  
However, Sobol' \gls{gsa} concerns scalar \glspl{qoi} only, hence, it is not suitable for multivariate \glspl{qoi} like electric machine efficiency maps and profiles. 
\textcolor{black}{
The same issue arises from alternative, non-variance-based \gls{gsa} methods, for example ones utilizing \gls{dgsm} \cite{kucherenko2016derivative, lamboni2020derivative}, Cram{\'e}r-von Mises distance \cite{gamboa2018sensitivity}, density-based indices \cite{borgonovo2007new}, or Shapley effects~\cite{song2016shapley}.
}

\textcolor{black}{
Considering \gls{gsa} for multivariate \glspl{qoi}, two main approaches exist. 
The first is to apply a \gls{gsa} method intended for scalar outputs elementwise, i.e., once per component of the \gls{qoi}. 
This results in a so-called sensitivity map per uncertain input parameter that has the same dimension as the \gls{qoi}, since one sensitivity index per \gls{qoi}-component is computed for each input~\cite{marrel2017sensitivity}.}
\textcolor{black}{
However, this elementwise approach runs into significant practical problems. 
First, the number of \gls{qoi}-components might render its application computationally demanding, if not prohibitive.
Second, even if the computational cost is acceptable, the large volume of generated data is difficult to be synthesized into a clear ranking of parameter importance~\cite{lamboni2009multivariate, lamboni2011multivariate}. 
This complexity hinders straightforward intepretation and subsequent decision-making, e.g., sensitivity-driven model simplification.
Third, the resulting sensitivity maps can be misleading, as sensitivity indices may become artificially inflated for output components with very small variance. 
It is therefore possible that importance is attributed to parameters with negligible absolute effect, entirely due to numerical artifacts. 
} 

\textcolor{black}{
The second approach is to utilize \gls{gsa} methods developed specifically for multivariate \glspl{qoi}.
This class of methods contains variance-based approaches that generalize Sobol' \gls{gsa} to multivariate \glspl{qoi}~\cite{campbell2006sensitivity, gamboa2013sensitivity}, as well as generalized sensitivity indices based on reproducing kernel Hilbert spaces \cite{barr2022generalized, barr2023kernel}, optimal transport theory \cite{borgonovo2025global, chiani2025global}, or \gls{dgsm} \cite{lamboni2021multivariate}. 
The main benefit of multivariate \gls{gsa} methods is that they provide a single generalized sensitivity index per uncertain input parameter, which quantifies its relative importance upon the full \gls{qoi}.}
\textcolor{black}{These generalized sensitivity metrics eliminate the need to interpret sensitivity maps, avoid the risk of numerical artifacts, and provide a clear ranking of the uncertain input parameters in terms of their influence on the multivariate \gls{qoi}.}

\textcolor{black}{
Within engineering design applications, it remains common practice to use scalar \gls{gsa} measures to multivariate \glspl{qoi} on an elementwise basis.
For the particular use-case of electric machine design, no relevant studies on leveraging multivariate \gls{gsa} methods exist, at least to the best knowledge of the authors.
This work aims to bridge this gap. 
We specifically focus on variance-based multivariate \gls{gsa} \cite{campbell2006sensitivity, gamboa2013sensitivity} and demonstrate its application to the model-based estimation of efficiency maps and profiles of a \gls{pmsm}.
In particular, an \gls{ecm} with uncertain circuit elements and an \gls{iga} model with geometry and material uncertainties are examined. 
For comparison purposes, we also employ the approach of applying Sobol' \gls{gsa} elementwise. 
The results showcase the limitations of the elementwise approach and highlight the advantages of the multivariate \gls{gsa} method.}
\textcolor{black}{
Most importantly, it is shown that multivariate \gls{gsa} avoids the pitfall of numerical artifacts that lead to misleading sensitivity information, which is in fact encountered when elementwise \gls{gsa} is employed.
Additionally, by providing a single (generalized) sensitivity index per parameter, multivariate \gls{gsa} allows to identify more accurately the relative impact of each uncertain input parameter on the full efficiency map or profile. 
This enables the simplification of the corresponding models by fixing non-influential parameters to their nominal values, while at the same time preserving \gls{uq} accuracy.
}

The rest of this paper is organized as follows. In Section~\ref{sec:eff} we recall the main principles behind the estimation of electric machine efficiency maps and profiles.
Section \ref{sec:sa} details variance-based \gls{gsa} for both scalar and multivariate \glspl{qoi}, including computation methods based on either \gls{mcs}~\cite{jansen1999analysis, saltelli2010variance, gamboa2016statistical, lamboni2011multivariate, gamboa2013sensitivity} or \glspl{pce}~\cite{sudret2008global, garcia2014global}.
Section \ref{sec:numexp} presents numerical studies based on an \gls{ecm} and an \gls{iga} model of a \gls{pmsm}.
The paper concludes with Section~\ref{sec:conclusion}, which summarizes the findings of this work and discusses follow-up research possibilities.

\section{Electric machine efficiency maps and profiles}
\label{sec:eff}
Electric machine efficiency, denoted with $\eta$, is defined as the ratio of output power $P_{\mathrm{out}}$ to input power $P_{\mathrm{in}}$, such that 
\begin{equation}
\eta = \frac{P_\mathrm{out}}{P_{\mathrm{in}}} = \frac{P_\mathrm{out}}{P_\mathrm{out} + P_{\mathrm{loss}}},
\end{equation}
where $P_{\mathrm{loss}}$ represents the total power dissipation due to heat, friction, or other losses \cite{roshandel2021losses}.
In motoring mode, mechanical power is the useful output, while electrical power is the input. 
Conversely, in generating mode, electrical power becomes the output, and mechanical power is the input. 
For \glspl{pmsm}, the output power is estimated as $P_{\mathrm{out}} = T \,\omega_{\mathrm{m}}$, where $T$ denotes the torque and $\omega_{\mathrm{m}}$ the angular velocity. 
The power loss term is commonly decomposed into three components, such that $P_{\mathrm{loss}} = P_{\mathrm{elec}} + P_{\mathrm{magn}} + P_{\mathrm{mech}}$,
where $P_{\mathrm{elec}}$ represents Joule losses in the copper windings of rotor and stator, $P_{\mathrm{magn}}$ corresponds to hysteresis and eddy current losses in the stator's iron core, and $P_{\mathrm{mech}}$ accounts for speed-dependent mechanical losses such as friction and windage \cite{krishnan2017permanent}.
The loss components exhibit complex dependencies to the operating points defined by specific $\left(T, \omega_\text{m}\right)$ pairs, hence $P_{\mathrm{loss}} = P_{\mathrm{loss}}(T, \omega_{\mathrm{m}})$. 
Naturally, these dependencies affect efficiency as well.

The efficiency map captures the machine's performance across its operating range by showing the maximum achievable efficiency at each torque-speed combination.
Aside from torque and speed values, the maximum efficiency is additionally dependent on operating conditions and constraints, such as current rating, voltage saturation, and thermal limits \cite{kahourzade2020estimation, hwang2021coupled}. 
In practice, producing an efficiency map entails: (i) discretizing the $T - \omega_{\text{m}}$ plane into a grid; (ii) computing the maximum efficiency upon that grid; (iii) interpolating the resulting values to create a continuous map.
The efficiency map is typically visualized as a contour plot over the torque-speed plane, bounded by an envelope that represents the machine's maximum torque capability across different speed ranges.
Efficiency profiles are produced similar to efficiency maps, with the difference that the $T - \omega_{\text{m}}$ grid is replaced by the operating points defined by the specific driving cycle \cite{dhakal2025comparative}. 

\section{Variance-based global sensitivity analysis}
\label{sec:sa}
Sensitivity analysis is broadly defined as ``the study of how uncertainty in the output of a model can be apportioned to different sources of uncertainty in the model input'' \cite{saltelli2004sensitivity}. 
There exist two main classes of sensitivity analysis methods: local methods that examine variations around a nominal point, and global methods that explore the entire input space \cite{saltelli2008global, razavi2021future}. 
Variance-based methods represent a prominent category within \gls{gsa} and are the focus of this work. 
The most well-known variance-based \gls{gsa} method is the Sobol' method \cite{sobol2001global}, which however concerns scalar outputs and can only be applied elementwise for multivariate outputs.
Multivariate \gls{gsa} methods were developed later \cite{campbell2006sensitivity, gamboa2013sensitivity}.
Sobol' and multivariate \gls{gsa}, as well as corresponding computation methods, are presented in more detail next.

In the following, we consider a parameter-dependent model  $f(\mathbf{x})$, such that $
f \colon \mathbf{x} \to \mathbf{y}$, equivalently, $\mathbf{y} = f\left(\mathbf{x}\right)$, 
where $\mathbf{x} \in \mathbb{R}^N$ is the input parameter vector and $\mathbf{y} \in \mathbb{R}^M$ the model output.
For $M=1$, the model has a scalar output $y \in \mathbb{R}$.
The model is assumed to be deterministic, however, the input parameters $\mathbf{x}$ are assumed to be realizations of a random vector $\mathbf{X}=(X_1,X_2,\dots,X_N)^\top$, where  $X_n,\,n=1,2,\dots,N$, are independent random variables.
The random vector is defined on the probability space $(\Theta,\Sigma,P)$, where $\Theta$ is the sample space, $\Sigma$ the sigma algebra of events, and $P \colon \Sigma \to [0,1]$ a probability measure \cite{smith2024uncertainty}.
For an outcome $\theta \in \Theta$, $\mathbf{x} = \mathbf{X}(\theta) \in \mathbb{R}^N$ is a random vector realization. 
The joint \gls{pdf} of $\mathbf{X}$ is denoted with $\varrho_{\mathbf{X}}(\mathbf{x})$, such that $\varrho_{\mathbf{X}} \colon \mathbb{R}^N \to \mathbb{R}_{\geq 0}$.
Due to the independence assumption, $\varrho_{\mathbf{X}}(\mathbf{x}) = \prod_{n=1}^N \varrho_{X_n}(x_n)$, where $\varrho_{X_n}(x_n)$ are the marginal (univariate) \glspl{pdf}.
Due to the propagation of uncertainty through the model, the output is a random vector dependent on the random input vector, such that $\mathbf{Y} = f(\mathbf{X})$ \cite{betz2014numerical}.
Note that input random fields can also be considered, in which case the \gls{kle} is typically employed to obtain discrete random inputs~\cite{betz2014numerical}.

\subsection{Sobol' sensitivity analysis}
\label{sec:sobol_sa}
Assuming a scalar output $Y = f(\mathbf{X})$ and a square-integrable $f(\mathbf{X})$, the output's variance $\operatorname{Var}\left(Y\right)$ can be decomposed as
\begin{equation}
\label{eq:var_decomposition}
\operatorname{Var}\left(Y\right) = \sum_{n=1}^N V_n + \sum_{n<k}^N V_{nk} + \cdots + V_{1,\ldots,N},
\end{equation}
where $V_n$ is the variance contribution of $X_n$ alone, $V_{nk}$ the contribution of $X_n$ interacting with $X_k$, and so on for higher-order interactions \cite{sobol2001global}.
Then, first-order Sobol' indices are given by 
\begin{equation}
\label{eq:first_order}
S_n = \frac{V_n}{\operatorname{Var}\left(Y\right)}=\frac{\operatorname{Var}_{X_n}\left( \mathbb{E}_{\mathbf{X}_{\sim n}}[Y \mid X_n] \right)}{\operatorname{Var}\left(Y\right)},
\end{equation} 
where $\operatorname{Var}_{X_n}\left( \mathbb{E}_{\mathbf{X}_{\sim n}}[Y \mid X_n] \right)$ is the first-order effect of $X_n$ and $\mathbf{X}_{\sim n}$ denotes the set of all variables except $X_n$.
$S_n$ quantifies the fractional contribution of $X_n$ alone to the output variance.  
Higher-order indices, for example, second-order indices $S_{nk} = V_{nk}/\operatorname{Var}(Y)$, quantify interaction effects. 
However, computing all $2^N - 1$ possible indices is often computationally challenging, if not prohibitive.
Total-order Sobol' indices provide a more practical alternative by capturing the total effect of $X_n$ including all interactions. 
The total effect of $X_n$ is $\mathbb{E}_{\mathbf{X}_{\sim n}}\left[ \operatorname{Var}_{X_n} \left( Y \mid \mathbf{X}_{\sim n} \right) \right]$ and the corresponding total-order Sobol' index is given by 
\begin{equation}
\label{eq:total_effect}
S_{\mathrm{T}n} = 1 - \frac{V_{\sim n}}{\operatorname{Var}(Y)}=\frac{\mathbb{E}_{\mathbf{X}_{\sim n}}\left[ \operatorname{Var}_{X_n} \left( Y \mid \mathbf{X}_{\sim n} \right) \right]}{\operatorname{Var}(Y)}.
\end{equation}
The difference $S_{\mathrm{T}n} - S_n$ quantifies the total interaction effect involving $X_n$. 
That is, if $S_{\mathrm{T}n} \gg S_n$, the variable $X_n$ participates in significant interactions with other parameters.
Conversely, if $S_{\mathrm{T}n} \approx S_n$, then $X_n$ has minimal interactions.

\subsection{Multivariate sensitivity analysis}
\label{sec:multivariate_sa}
Assuming a multivariate output  $\mathbf{Y} = f(\mathbf{X})$ with square-integrable $f(\mathbf{X})$, the covariance matrix $\mathbf{C} = \operatorname{Cov}(\mathbf{Y})$ can be decomposed analogously to the (scalar) variance decomposition \eqref{eq:var_decomposition}, such that 
\begin{equation}
\label{eq:covar_decomposition}
\mathbf{C} = \sum_{n=1}^{N} \mathbf{C}_n + \sum_{n<k}^N \mathbf{C}_{nk} + \cdots + \mathbf{C}_{1,\ldots,N},
\end{equation}
where each matrix $\mathbf{C}_n$, $\mathbf{C}_{nk}$, etc., represents the covariance contribution from the corresponding subset of input variables~\cite{gamboa2013sensitivity, garcia2014global}. 
This decomposition captures both main (first-order) effects and interactions across all components of the multivariate output.
The trace $\mathrm{tr}\left(\mathbf{C}\right)$ corresponds to summing the variances across all output components and is used to obtain scalar sensitivity indices from the covariance decomposition. 
Then, the first-order generalized sensitivity index for input $X_n$ is defined as
\begin{equation}
\label{eq:GSn}
G_n = \frac{\operatorname{tr}(\mathbf{C}_n)}{\operatorname{tr}(\mathbf{C})} = \frac{\sum_{m=1}^M V_n^{(m)}}{\sum_{m=1}^M \operatorname{Var}\left(Y_m\right)},
\end{equation}
where $V_n^{(m)}$ quantifies the fractional contribution of $X_n$ alone to the variance of the $m$-th component of the $M$-dimensional output.
Summing over all output components quantifies the proportion of the total output variance attributable solely to $X_n$.
Accordingly, the total-order generalized sensitivity index is given by
\begin{equation}
\label{eq:GSTn}
G_{\mathrm{T}n} = 1 - \frac{\operatorname{tr}(\mathbf{C}_{\sim n})}{\operatorname{tr}(\mathbf{C})} = 1 - \frac{\sum_{m=1}^M V_{\sim n}^{(m)}}{\sum_{m=1}^M \operatorname{Var}\left(Y_m\right)},
\end{equation}
and captures all effects involving $X_n$, including its interactions.
The difference $G_{\mathrm{T}n} - G_n$ quantifies the total interaction effect of $X_n$.
The generalized indices provide a comprehensive assessment of parameter importance for multivariate outputs, while avoiding the interpretability issues associated with the elementwise application of Sobol' \gls{gsa}~\cite{lamboni2011multivariate}.

\textcolor{black}{\subsection{Computation of sensitivity indices}}
\label{sec:sa_computation}
The Sobol' and generalized sensitivity indices defined in Sections \ref{sec:sobol_sa} and \ref{sec:multivariate_sa} can be computed using various numerical techniques. 
In this work, sensitivity indices are computed using either \gls{mcs} or \gls{pce}. 
The corresponding computations are described in Sections \ref{sec:mc_sa} and \ref{sec:pce-sa}, respectively.

\subsubsection{Computation based on Monte Carlo sampling}
\label{sec:mc_sa}
A standard approach for estimating Sobol' indices is the \gls{mcs}-based ``pick-and-freeze'' method \cite{jansen1999analysis, saltelli2010variance, gamboa2016statistical}, which is also easily generalized to multivariate outputs \cite{lamboni2011multivariate, gamboa2013sensitivity}.
The method involves generating random samples of the input vector $\mathbf{X}$ according to its joint probability distribution and then constructing specific sample sets where one parameter is ``frozen'' while the rest are re-sampled.

We consider the scalar output case first. 
Given two independent sampling matrices $\mathbf{A}$ and $\mathbf{B}$ of size $N_{\mathrm{s}} \times N$, where $N_{\mathrm{s}}$ is the number of samples, the hybrid matrices $\mathbf{A}_B^{(n)}$ are constructed by replacing the $n$-th column of $\mathbf{A}$ with the $n$-th column from $\mathbf{B}$. 
Evaluating the model on the $j$-th row of these matrices yields scalar output samples $\mathbf{y}_{\mathbf{A},j} = f(\mathbf{A})_j$, $\mathbf{y}_{\mathbf{B},j} = f(\mathbf{B})_j$, and $\mathbf{y}_{\mathbf{A}_B^{(n)},j} = f\left(\mathbf{A}_B^{(n)}\right)_j$, $j=1,\dots,N_{\mathrm{s}}$, where $\mathbf{y}_{\mathbf{A}} = f(\mathbf{A})$, $\mathbf{y}_{\mathbf{B}} = f(\mathbf{B})$, $\mathbf{y}_{\mathbf{A}_B^{(n)}} = f\left(\mathbf{A}_B^{(n)}\right)$ are $N_\mathrm{s}$-dimensional vectors. 
The main and total effects (see Section~\ref{sec:sobol_sa}) can then be estimated as 
\begin{subequations} 
\begin{align}
V_{n} &\approx \frac{1}{N_{\mathrm{s}}} \sum_{j=1}^{N_{\mathrm{s}}} f\left(\mathbf{B}\right)_j \left( f\left(\mathbf{A}_B^{(n)}\right)_j - f\left(\mathbf{A}\right)_j\right), \\
V_{\sim n} &\approx \frac{1}{2N_{\mathrm{s}}} \sum_{j=1}^{N_{\mathrm{s}}} \left( f\left(\mathbf{A}\right)_j - f\left(\mathbf{A}_B^{(n)}\right)_j\right)^2,
\end{align}
\end{subequations}
while the variance estimation is given by
\begin{equation}
\operatorname{Var}(Y) \approx \frac{1}{N_{\mathrm{s}}} \sum_{j=1}^{N_{\mathrm{s}}} f(\mathbf{A})_j^2 - \left( \frac{1}{N_{\mathrm{s}}} \sum_{j=1}^{N_{\mathrm{s}}} f(\mathbf{A})_j \right)^2,
\end{equation}
Using these estimates within \eqref{eq:first_order} and \eqref{eq:total_effect} leads to the pick-and-freeze estimates of the first- and total-order Sobol' indices.

Moving on to multivariate outputs, the matrices $\mathbf{A}$, $\mathbf{B}$, and $\mathbf{A}_B^{(n)}$ remain as in the scalar case and model evaluations on the $j$-th row of these matrices now result in $M$-dimensional output samples $\mathbf{Y}_{\mathbf{A},j}= f\left(\mathbf{A}\right)_j$, $\mathbf{Y}_{\mathbf{B},j} = f\left(\mathbf{B}\right)_j$, $\mathbf{Y}_{\mathbf{A}_B^{(n)},j}=f\left(\mathbf{A}_B^{(n)}\right)_j$, $j=1,\dots,N_{\mathrm{s}}$, where $\mathbf{Y}_{\mathbf{A}}= f\left(\mathbf{A}\right)$, $\mathbf{Y}_{\mathbf{B}} = f\left(\mathbf{B}\right)$, $\mathbf{Y}_{\mathbf{A}_B^{(n)}}= f\left(\mathbf{A}_B^{(n)}\right)$ are matrices of size $N_{\mathrm{s}} \times M$.
Then, the output covariance matrix can be estimated by
\begin{align}
\mathbf{C} \approx &\frac{1}{N_{\mathrm{s}}} \sum_{j=1}^{N_{\mathrm{s}}}  \left(f(\mathbf{A})_j \right) \left(  f(\mathbf{A})_j \right)^\top \nonumber \\  
&- \left(\frac{1}{N_{\mathrm{s}}} \sum_{j=1}^{N_{\mathrm{s}}} f\left(\mathbf{A}\right)_j\right) \left(\frac{1}{N_{\mathrm{s}}} \sum_{j=1}^{N_{\mathrm{s}}} f\left(\mathbf{A}\right)_j\right)^\top,
\end{align}
while the estimates for the covariance matrices $\mathbf{C}_n$ and $\mathbf{C}_{\sim n}$ are given as
\begin{subequations}
\begin{align}
\mathbf{C}_n \approx &\frac{1}{N_{\mathrm{s}}} \sum_{j=1}^{N_{\mathrm{s}}} f\left(\mathbf{B}\right)_j f\left(\mathbf{A}_B^{(n)}\right)_j^\top \nonumber \\ 
&- \left(\frac{1}{N_{\mathrm{s}}} \sum_{j=1}^{N_{\mathrm{s}}} f\left(\mathbf{A}\right)_j\right) \left(\frac{1}{N_{\mathrm{s}}} \sum_{j=1}^{N_{\mathrm{s}}} f\left(\mathbf{B}\right)_j\right)^\top,
\end{align}
\begin{align}
\mathbf{C}_{\sim n} \approx \frac{1}{2 N_{\mathrm{s}}} \sum_{j=1}^{N_{\mathrm{s}}} \Biggl[ &\left( f\left(\mathbf{A}\right)_j - f\left(\mathbf{A}_B^{(n)}\right)_j \right) \nonumber \\ 
&\left( f\left(\mathbf{A}\right)_j- f\left(\mathbf{A}_B^{(n)}\right)_j \right)^\top \Biggr].
\end{align}
\end{subequations}
Using these estimates in \eqref{eq:GSn} and \eqref{eq:GSTn} leads to the pick-and-freeze estimates for the first- and total-order generalized sensitivity indices.

\subsubsection{Computation based on polynomial chaos expansion}
\label{sec:pce-sa}
The \gls{pce} is a polynomial approximation of the form
\begin{equation}
 \label{eq:pce-truncated}
\mathbf{y} = f \left(\mathbf{x} \right) \approx \widehat{f} \left(\mathbf{x} \right) = \sum_{\boldsymbol{\alpha} \in \Lambda} \mathbf{c}_{\boldsymbol{\alpha}} \Psi_{\boldsymbol{\alpha}} \left( \mathbf{x} \right), 
\end{equation}
where $\left\{ \Psi_{\boldsymbol{\alpha}}\left(\mathbf{x}\right)  \right\}_{\boldsymbol{\alpha} \in \Lambda}$ is a multivariate polynomial basis which is orthonormal with respect to the joint \gls{pdf} of the input random vector $\mathbf{X}$, $\boldsymbol{\alpha} = \left(\alpha_1, \dots \alpha_N\right)$ is a multi-index holding the polynomial degrees of the univariate polynomials $\psi_n^{(\alpha_n)}\left(x_n\right)$ that form the multivariate polynomials as $\Psi_{\boldsymbol{\alpha}}\left(\mathbf{x}\right) = \prod_{n=1}^N \psi_n^{(\alpha_n)} \left(x_n\right)$, $\Lambda$ is a multi-index set containing all multi-indices that define the multivariate polynomial basis, and $\mathbf{c}_{\boldsymbol{\alpha}} \in \mathbb{R}^M$ are coefficients with the same dimensions as the output \cite{xiu2002wiener}. 
Exploiting the orthonormality of the \gls{pce} basis, it is straightforward to show that
\begin{subequations}
\begin{align}
\mathbb{E}\left[\mathbf{Y}\right] &\approx \mathbf{c}_\mathbf{0}, \\
\operatorname{Var}\left(\mathbf{Y}\right) &\approx \sum_{\boldsymbol{\alpha} \in \Lambda \setminus \mathbf{0}} \mathbf{c}_{\boldsymbol{\alpha}}^2,
\end{align}
\end{subequations}
where the zeroth multi-index is denoted as $\mathbf{0} = \left(0,0,\dots,0\right)$.

A total-degree basis is commonly used, in which case $\Lambda = \left\{ \boldsymbol{\alpha}: \left| \boldsymbol{\alpha} \right|_1 \leq P \right\}$, where $P$ denotes the maximum polynomial degree.
The size of the total-degree basis is equal to $\#\Lambda = \frac{\left(N+P\right)!}{N! P!}$. 
To mitigate computational issues arising for increasing $N$, $P$, various methods for constructing sparse bases have been suggested \cite{luethen2021sparse, luethen2022automatic}.

The \gls{pce} coefficients are computed here using a regression-based approach \cite{hadigol2018least}. 
Alternative methods like pseudo-spectral projection \cite{constantine2012sparse, conrad2013adaptive} or interpolation \cite{buzzard2012global, galetzka2023hp} are not considered in this work.
For the regression-based coefficient estimation, we assume a set of input parameter realizations, commonly called the experimental design, and the corresponding model responses, i.e., $\left\{\mathbf{x}^{(j)}, \mathbf{y}^{(j)}\right\}_{j=1}^{N_{\mathrm{s}}}$. 
The coefficients are then computed by solving the least squares minimization problem
\begin{equation} 
\label{eq:regression-matrix}
\underset {\left(\mathbf{c}_{\boldsymbol{\alpha}}\right)_{\boldsymbol{\alpha} \in \Lambda}}{\min} 
\left\{\sum_{j=1}^{N_{\mathrm{s}}}\left( \mathbf{y}^{(j)} - \sum_{\boldsymbol{\alpha} \in \Lambda} \mathbf{c}_{\boldsymbol{\alpha}} \Psi_{\boldsymbol{\alpha}}\left(\mathbf{x}^{(j)}\right)\right)^2 \right\}.
\end{equation}
To ensure a well determined least squares problem, it is common to use an oversampling coefficient $C \geq 1$, such that $N_{\mathrm{s}} = C \, \#\Lambda$, where $\#$ denotes the cardinality of the multi-index set, equivalently, the size of the polynomial basis. 
The value of the oversampling coefficient is crucial for the stability of the least squares problem \eqref{eq:regression-matrix} and the accuracy of the \gls{pce}~\cite{migliorati2013approximation, migliorati2014analysis}.

First assuming a scalar model output, it is straightforward to show that the squared (scalar) coefficients $c_{\boldsymbol{\alpha}}^2$, $\boldsymbol{\alpha} \ne \mathbf{0}$, are equivalent to the partial variances that appear in the variance decomposition \eqref{eq:var_decomposition} \cite{sudret2008global}.
This allows to compute Sobol' indices by simply post-processing the \gls{pce}.
To that end, the multi-indices corresponding to partial variances caused by $X_n$, either individually (main  effect) or in combination with all other random variables (total effect), are collected into the multi-index sets $\Lambda_n \subset \Lambda$ and $\Lambda_{\mathrm{T}n} \subset \Lambda$, respectively defined as
\begin{subequations}
	\begin{align}
		\Lambda_n &= \{\boldsymbol{\alpha} \in \Lambda \; : \; \alpha_n \neq 0 \:\: \text{and} \:\: \alpha_l = 0, l = 1,\dots,N, l \neq n\}, \\
		\Lambda_{\text{T}n} &= \{\boldsymbol{\alpha} \in \Lambda \; : \; \alpha_n \neq 0\}.
	\end{align}
\end{subequations}
The first- and total-effect Sobol' indices are then estimated as 
\begin{subequations}
\label{eq:pce-sobol}
\begin{align}
    S_n &\approx \frac{\sum_{\boldsymbol{\alpha} \in \Lambda_n} c_{\boldsymbol{\alpha}}^2}{\sum_{\boldsymbol{\alpha} \in \Lambda \setminus \mathbf{0}} c_{\boldsymbol{\alpha}}^2}, \\
    S_{\mathrm{T}n} &\approx \frac{\sum_{\boldsymbol{\alpha} \in \Lambda_{\mathrm{T}n}} c_{\boldsymbol{\alpha}}^2}{\sum_{\boldsymbol{\alpha} \in \Lambda \setminus \mathbf{0}} c_{\boldsymbol{\alpha}}^2}.
\end{align}
\end{subequations}

The approach is similar for multivariate outputs, where we leverage the fact that the traces of the covariance matrices in \eqref{eq:covar_decomposition} are equal to the sum of the variances of all response components, dependent on the specific input random variable combinations \cite{garcia2014global}.
Then, the first- and total-order generalised sensitivity indices can be estimated as 
\begin{subequations}
\label{eq:pce-gen-sobol}
\begin{align}
G_n &\approx \frac{\sum_{m=1}^M \left(\sum_{\boldsymbol{\alpha} \in \Lambda_n} c_{m,\boldsymbol{\alpha}}^2\right)}{\sum_{m=1}^M \left(\sum_{\boldsymbol{\alpha} \in \Lambda \setminus \mathbf{0}} c_{m,\boldsymbol{\alpha}}^2\right)},\\
G_{\mathrm{T}n} &\approx \frac{\sum_{m=1}^M \left(\sum_{\boldsymbol{\alpha} \in \Lambda_{\mathrm{T}n}} c_{m,\boldsymbol{\alpha}}^2\right)}{\sum_{m=1}^M \left(\sum_{\boldsymbol{\alpha} \in \Lambda \setminus \mathbf{0}} c_{m,\boldsymbol{\alpha}}^2\right)},
\end{align}
\end{subequations}
where $\mathbf{c}_{\boldsymbol{\alpha}} = \left(c_{1, \boldsymbol{\alpha}}, \dots, c_{M, \boldsymbol{\alpha}}\right)^\top$.

\subsection{Computational cost}
\label{sec:sa_cost}
\textcolor{black}{
Using either \gls{mcs} or \gls{pce} to compute the Sobol' or generalized sensitivity indices, the \gls{qoi} must be sampled $N_{\text{s}}$ times for different realizations of the uncertain model parameters. 
As shown in Section~\ref{sec:mc_sa}, \gls{mcs}-based \gls{gsa} requires additional model sampling on input combinations formed from the original sample. 
The total sampling cost rises to $N_{\text{GSA}}^{\text{MCS}} = (N+2) N_{\text{s}}$. 
It is common for \gls{mcs} that $N_{\text{s}}$ is in the order of $10^3-10^5$, possibly even higher for sufficient convergence~\cite{saltelli2008global}. 
This can result in an undesirable computational cost, especially if the sampled model is computationally expensive.
}

\textcolor{black}{
For \gls{pce}-based \gls{gsa}, the sample is used to compute the \gls{pce} coefficients by solving the least-squares minimization problem \eqref{eq:regression-matrix}.
Hence, the sampling cost of \gls{pce}-based \gls{gsa} is  $N_{\text{GSA}}^{\text{PCE}} = N_{\text{s}}$. 
Nonetheless, the computational cost can still become intractable. 
As discussed in Section~\ref{sec:pce-sa}, a total-degree \gls{pce} basis grows rapidly for increasing input dimensions $N$ or maximum polynomial degree $P$}\textcolor{black}{, as $\#\Lambda = \frac{\left(N+P\right)!}{N! P!}$.
Additionally considering that $N_\text{s} = C \#\Lambda$, where the oversampling coefficient $C \geq 1$ is commonly chosen in the range $C \in \left[2,10\right]$ depending on the desired coefficient estimation accuracy and robustness \cite{migliorati2014analysis}, it becomes obvious that the use of total-degree \glspl{pce} is limited to low $P$, $N$.
For instance, with $N=10$ uncertain input parameters and a moderate polynomial degree $P=4$, the total-degree basis contains $\#\Lambda=1001$ terms. 
Using an oversampling coefficient $C=2$, which is a common choice for well-behaved models, results in $N_s=2002$ samples. 
Using $C=5$ for more robust coefficient estimation, the number of samples rises to $N_s=5005$. 
Such sample sizes may be acceptable for computationally inexpensive models, but can become prohibitive for computationally demanding model evaluations. 
}

\textcolor{black}{
The rapid growth in \gls{pce} complexity with increasing input dimensions is a manifestation of the so-called ``curse of dimensionality''. 
This issue is commonly mitigated with the use of sparse \glspl{pce} \cite{luethen2021sparse, luethen2022automatic}, for example ones based on adaptive basis construction algorithms \cite{loukrezis2020robust, loukrezis2025multivariate}. 
These methods retain the most significant basis terms and omit the rest, thus reducing basis size without sacrificing approximation accuracy, provided that the model $f(\mathbf{x})$ exhibits sufficient sparsity in its polynomial representation. 
Besides sparsity, the extent of the usability of sparse/adaptive \glspl{pce} is also connected to the regularity of the problem at hand.
In settings where $f(\mathbf{x})$ exhibits reduced regularity, e.g., sharp gradients are present, higher polynomial degrees are needed to reach the desired approximation accuracy, which further exacerbates the curse of dimensionality. 
Reduced regularity is typically addressed with dedicated multi-element methods that employ some sort of local refinement during \gls{pce} construction \cite{galetzka2023hp, basmaji2022anisotropic}.
Under strict assumptions on regularity and sparsity, it has been shown that sparse/adaptive \glspl{pce} can overcome the curse of dimensionality in a theoretical sense \cite{cohen2018multivariate}.
However, these assumptions rarely hold in practically relevant problem settings such as engineering applications. 
Practical experience indicates that sparse/adaptive \glspl{pce} remain usable typically up to moderately high input dimensions, possibly up to several tens.
However, the extent of their usability is highly problem-dependent.
Last, very high dimensional \glspl{qoi} can also lead to increased computational cost, as they require the least squares problem \eqref{eq:regression-matrix} to be solved a number of times equal to the dimension of the \gls{qoi}.
This issue can in most cases be effectively countered with the use of suitable dimension reduction methods \cite{hou2022dimensionality}.
}

\textcolor{black}{
In summary, \glspl{pce} can be much more computationally efficient than \gls{mcs} for relatively smooth models with up to moderately high input dimensions. 
Total-degree \glspl{pce} are suitable for problems where the number of uncertain parameters is modest and the required polynomial degree is low, or if the model evaluation cost is sufficiently low.
Sparse/adaptive \gls{pce} methods should be employed for problems with moderately high input dimensions, assuming sufficient model sparsity.
Multi-element and locally adaptive \gls{pce} methods should be selected for models with reduced regularity.
Nonetheless, there exist settings where \gls{pce} may be altogether not applicable, e.g., for strongly non-smooth and very high-dimensional models. 
In those cases, \gls{mcs} should be selected, as its convergence rate is independent of input dimensionality or regularity. 
However, this constant convergence rate is also slow (i.e., $\mathcal{O}\left( N_{\text{s}}^{-0.5} \right)$ in root mean square error), resulting in the aforementioned sampling cost.
}

\textcolor{black}{
Note that the cost discussion must be revisited for multivariate outputs like efficiency maps or profiles. 
In that case, the model must also be sampled on the operating points $(T, \omega_\text{m})$ that define the map or profile. 
The number of operating points, denoted with $N_\text{op}$, becomes an additional factor concerning the sample cost of \gls{gsa}, which increases to $N_{\text{GSA}}^{\text{MCS}} = (N+2) N_{\text{s}} N_{\text{op}}$ and $N_{\text{GSA}}^{\text{PCE}} = N_{\text{s}} N_{\text{op}}$, respectively.
}

\section{Application to permanent magnet synchronous machine models}
\label{sec:numexp}

As test-case, we consider the benchmark \gls{pmsm} from the work of Muetze et al. \cite{muetze2025creator}. 
In Section~\ref{sec:numexp-pmsm-ecm}, the \gls{ecm} with four uncertain circuit parameters is examined, first regarding its efficiency map, and second regarding its efficiency profile for the \gls{wltp} \cite{mock2014wltp} driving cycle.
In Section~\ref{sec:numexp-pmsm-iga}, a two-dimensional \gls{iga} model with eleven uncertain geometric and material parameters is examined. 
\textcolor{black}{Due to the computational cost of the \gls{iga} model, \gls{gsa} concerns only the efficiency map in motoring mode.}

\subsection{Equivalent circuit model}
\label{sec:numexp-pmsm-ecm}
\begin{figure}[t!]
\centering
\begin{subfigure}[b]{1\columnwidth}
    \centering
    \resizebox{0.7\columnwidth}{!}{


\begin{circuitikz}

  \draw  (0,0) -- (6.5,0) ;

  \draw[->, >=latex] (0,0.15) -- (0,2.85) node[midway,right] {$V_{\mathrm{p}}$};

  \draw (0,3) to[short, i_=$I_{\mathrm{p}}$, pos=0.5] (1,3);

  \draw (1,3) to[R, l={$R_{\mathrm{s}}$}] (3.5,3)
             to[L, l={$L_{\mathrm{s}}$}] (6,3);
  \draw (6,3) to[short] (6.5,3);
  \draw (6.5,3) to[sV] (6.5,0);

\end{circuitikz}

}
    \caption{Single-phase ECM.}
\end{subfigure}
\hfill
\begin{subfigure}[b]{1\columnwidth}
    \centering
    \resizebox{0.7\columnwidth}{!}{


\begin{circuitikz}

  \draw  (0,0) -- (6.5,0) -- (6.5,3);

  \draw[->, >=latex] (0,0.15) -- (0,2.85) node[midway,right] {$V_{\mathrm{d}}$};

  \draw (0,3) to[short, i_=$I_{\mathrm{d}}$, pos=0.5] (1,3);

  \draw (1,3) to[R, l={$R_{\mathrm{s}}$}] (3.5,3)
               to[L, l={$L_\text{d}$}] (6,3);

  \draw (6,3) to[short] (6.5,3);

  \draw (6.5,1.5) node[circle, draw=black, minimum size=0.9cm,fill=white, draw opacity=1,inner sep=1pt] {
    \begin{tabular}{@{}c@{}}
      $-$ \\[-2pt] $+$
    \end{tabular}
  };

\end{circuitikz}

}
    \caption{Direct axis (d-axis) ECM.}
\end{subfigure}
\hfill 
\begin{subfigure}[b]{1\columnwidth}
    \centering
    \resizebox{0.7\columnwidth}{!}{


\begin{circuitikz}

  \draw  (0,0) -- (6.5,0) -- (6.5,3);

  \draw[->, >=latex] (0,0.15) -- (0,2.85) node[midway,right] {$V_{\mathrm{q}}$};

  \draw (0,3) to[short, i_=$I_{\mathrm{q}}$, pos=0.5] (1,3);

  \draw (1,3) to[R, l={$R_{\mathrm{s}}$}] (3,3)
               to[L, l={$L_\text{q}$}] (5,3);

  \draw (5,3) to[short] (6.5,3);
  \draw (5.5,3) node[circle, draw=black, minimum size=0.9cm, fill=white, draw opacity=1,inner sep=1pt] {
  \begin{tabular}{@{}c@{}c@{}}
    $+$ & \raisebox{-0.3ex}{\rotatebox{90}{\raisebox{-2ex}{$-$}}}
  \end{tabular}
  };  

  \draw (6.5,1.5) node[circle, draw=black, minimum size=0.9cm,fill=white, draw opacity=1,inner sep=1pt] {
    \begin{tabular}{@{}c@{}}
      $+$ \\[-2pt] $-$
    \end{tabular}
  };

\end{circuitikz}

}
    \caption{Quadrature axis (q-axis) ECM.}
\end{subfigure}
\caption{\gls{pmsm} \gls{ecm}.}
\label{fig:pmsm-ecm}
\end{figure}
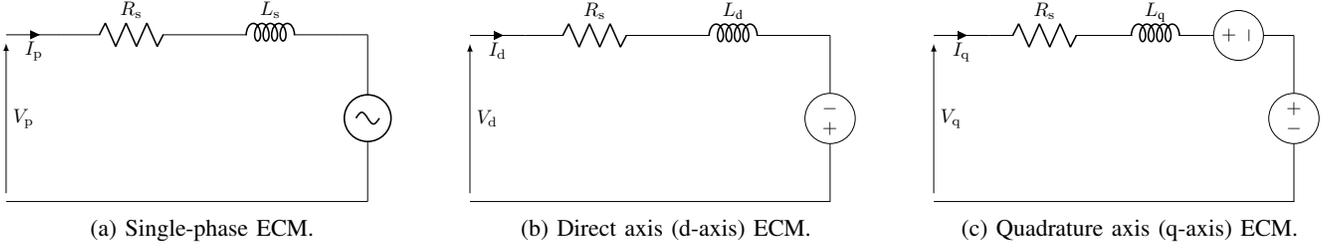

The \gls{ecm} is depicted in Figure~\ref{fig:pmsm-ecm}.
The corresponding semi-analytical computational model is based on the standard voltage and flux-linkage equations in the d-q frame and is available at a dedicated online repository \cite{dhakal_2024_sns1d-77m43}. 
Efficiency for a given operating point $(T, \omega_{\text{m}})$ can be estimated in a few milliseconds.
The four circuit elements, listed in Table~\ref{tab:PMSM-parameters}, are treated as uncertain and assumed to vary uniformly within  $\pm$5\% of their nominal values.

\begin{table}[t!]
\centering
\caption{Parameters of \gls{pmsm} \gls{ecm}.}
\label{tab:PMSM-parameters}
\begin{tabular}{lccc}
\toprule
Parameter & Symbol & Nominal value & Units \\
\midrule
Phase resistance & $R_{\mathrm{s}}$ & $8.9462$ & $\Omega$ \\
Magnet flux linkage & $\lambda$ & $0.1144$ & $\mathrm{Wb}$ \\
d-axis inductance & $L_\text{d}$ & $0.2055$ & $\mathrm{H}$ \\
q-axis inductance & $L_\text{q}$ & $0.332$ & $\mathrm{H}$ \\
\bottomrule
\end{tabular}
\end{table}

\subsubsection{Sensitivity analysis of efficiency map}
\label{sec:numexp-pmsm-ecm-grid}
\begin{figure}[t!]
\centering
\begin{subfigure}[b]{0.35\textwidth}
    \centering
    \includegraphics[width=\textwidth]{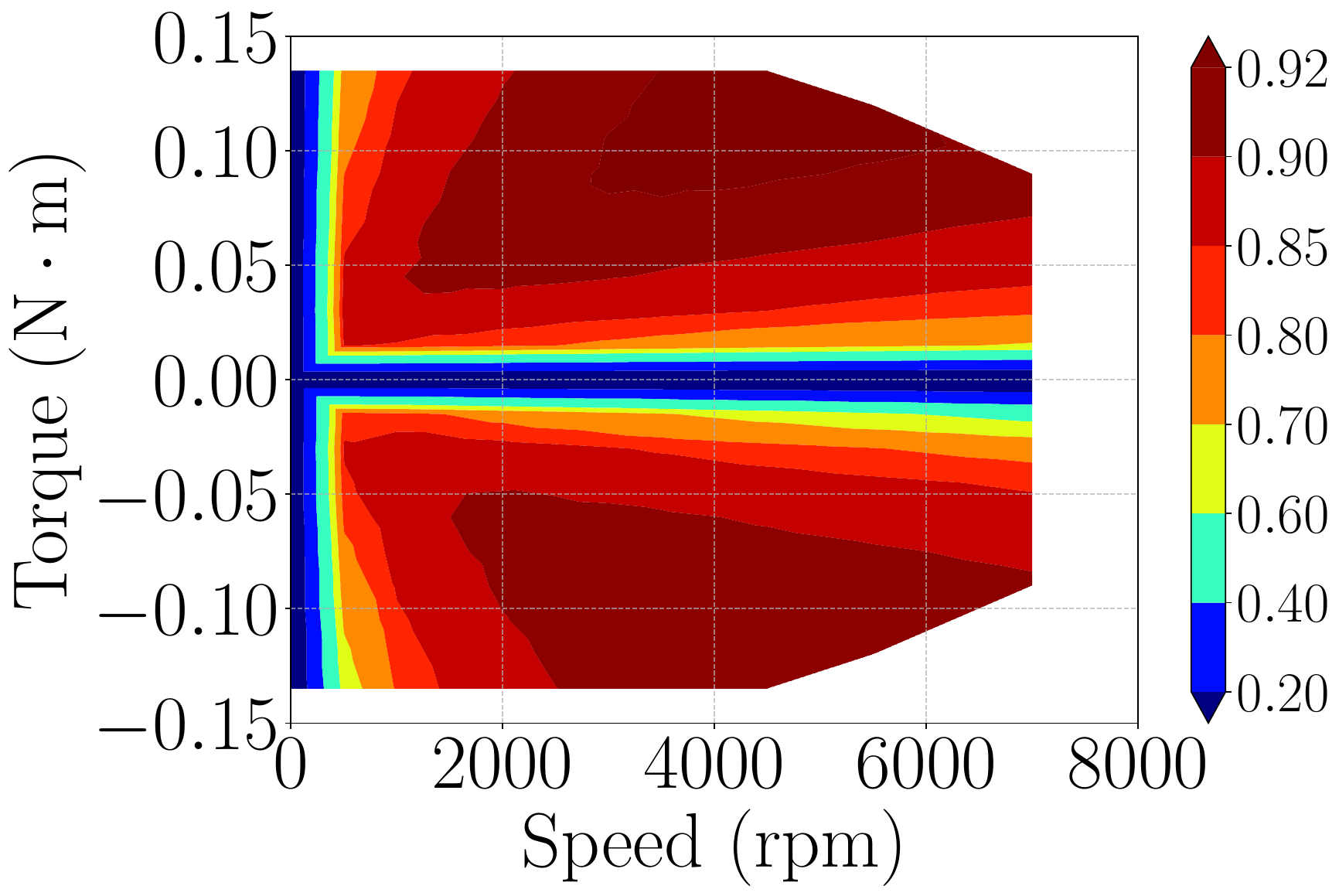}
    \caption{Mean.}
    \label{fig:pmsm-grid-uq-mean}
\end{subfigure}
\\
\begin{subfigure}[b]{0.35\textwidth}
    \centering
    \includegraphics[width=\textwidth]{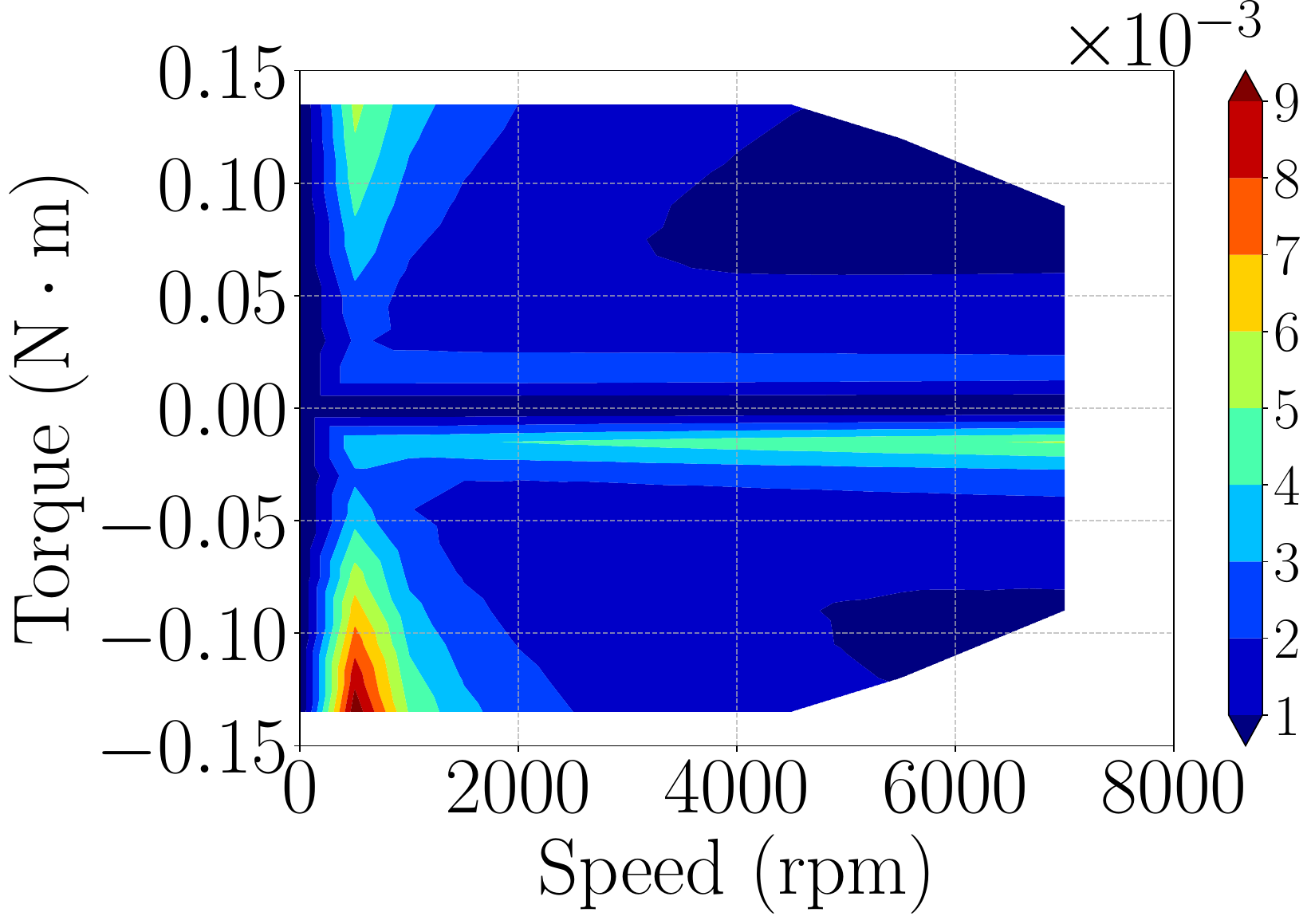}
    \caption{Standard deviation.}
    \label{fig:pmsm-grid-uq-std}
\end{subfigure}
\caption{\textcolor{black}{Mean and standard deviation of \gls{ecm}'s efficiency map.}}
\label{fig:pmsm-grid-uq}
\end{figure}

The efficiency map is evaluated on a grid of $N_{\text{op}} = 232$ operating points $\left(T, \omega_{\text{m}}\right)$, uniformly distributed over the operating range.
Figure~\ref{fig:pmsm-grid-uq} shows the mean and standard deviation of the efficiency map, computed with \gls{mcs} and a sample of $N_{\text{s}}^{\text{MCS}} = 2 \cdot 10^5$ model evaluations. 

\textcolor{black}{
Nearly identical results are obtained with a total-degree \gls{pce} with maximum polynomial degree $P=2$ and oversampling coefficient $C=2$, which corresponds to a sample size $N_\text{s}^{\text{PCE}} = 30$. 
These values for $P$ and $C$ have been selected over a range of possible combinations. 
Tables~\ref{tab:mean_eff_error} and \ref{tab:std_eff_error} show the \glspl{mae} in the mean and standard deviation of the efficiency map for total-degree \glspl{pce} with varying $P$ and $C$.
As can be observed, no significant error improvement is obtained by increasing $P$ or $C$ beyond the selected values. 
}

\begin{table}[t!]
\centering
\caption{\textcolor{black}{\Glspl{mae} in \gls{ecm}'s mean efficiency map for \glspl{pce}  with varying polynomial degree $P$ and oversampling coefficient $C$.}}
\label{tab:mean_eff_error}
\begin{tabular}{ccccc}
\toprule 
$P$ & $C=2$ & $C=3$ & $C=4$ & $C=5$ \\
\midrule 
$1$ & $7.69\times10^{-6}$ & $5.49\times10^{-6}$ & $6.80\times10^{-6}$ & $6.44\times10^{-6}$ \\
$2$ & $4.34\times10^{-6}$ & $4.30\times10^{-6}$ & $4.28\times10^{-6}$ & $4.28\times10^{-6}$ \\
$3$ & $4.27\times10^{-6}$ & $4.27\times10^{-6}$ & $4.27\times10^{-6}$ & $4.27\times10^{-6}$ \\
$4$ & $4.27\times10^{-6}$ & $4.27\times10^{-6}$ & $4.27\times10^{-6}$ & $4.27\times10^{-6}$ \\
$5$ & $4.27\times10^{-6}$ & $4.27\times10^{-6}$ & $4.27\times10^{-6}$ & $4.27\times10^{-6}$ \\
\bottomrule
\end{tabular}
\end{table}

\begin{table}[t!]
\centering
\caption{\textcolor{black}{\Glspl{mae} in standard deviation of \gls{ecm}'s efficiency map for \glspl{pce}  with varying polynomial degree $P$ and oversampling coefficient $C$.}}
\label{tab:std_eff_error}
\begin{tabular}{ccccc}
\toprule 
$P$ & $C=2$ & $C=3$ & $C=4$ & $C=5$ \\
\midrule 
$1$ & $6.97\times10^{-6}$ & $8.01\times10^{-6}$ & $7.44\times10^{-6}$ & $6.04\times10^{-6}$ \\
$2$ & $4.45\times10^{-6}$ & $4.49\times10^{-6}$ & $4.50\times10^{-6}$ & $4.48\times10^{-6}$ \\
$3$ & $4.46\times10^{-6}$ & $4.46\times10^{-6}$ & $4.46\times10^{-6}$ & $4.46\times10^{-6}$ \\
$4$ & $4.46\times10^{-6}$ & $4.46\times10^{-6}$ & $4.46\times10^{-6}$ & $4.46\times10^{-6}$ \\
$5$ & $4.46\times10^{-6}$ & $4.46\times10^{-6}$ & $4.46\times10^{-6}$ & $4.46\times10^{-6}$ \\
\bottomrule
\end{tabular}
\end{table}

Similar errors are obtained for the \gls{gsa} results, therefore, only one set of results is presented in the following. 
For the aforementioned sample sizes, the \gls{mcs}-based \gls{gsa} requires
$N_{\text{GSA}}^{\text{MCS}} = N_{\text{s}}^{\text{MCS}} \left(5+2\right) N_{\text{op}} = 3.25 \cdot 10^{8}$ model evaluations. 
The \gls{pce}-based approach needs
$N_{\text{GSA}}^{\text{PCE}} = N_{\text{s}}^{\text{PCE}} N_{\text{op}} = 6.96 \cdot 10^{3}$ model evaluations, further highlighting its advantage in terms of computational efficiency.

\begin{figure}[t!]
\centering
\begin{subfigure}[b]{0.35\textwidth}
    \centering
    \includegraphics[width=\textwidth]{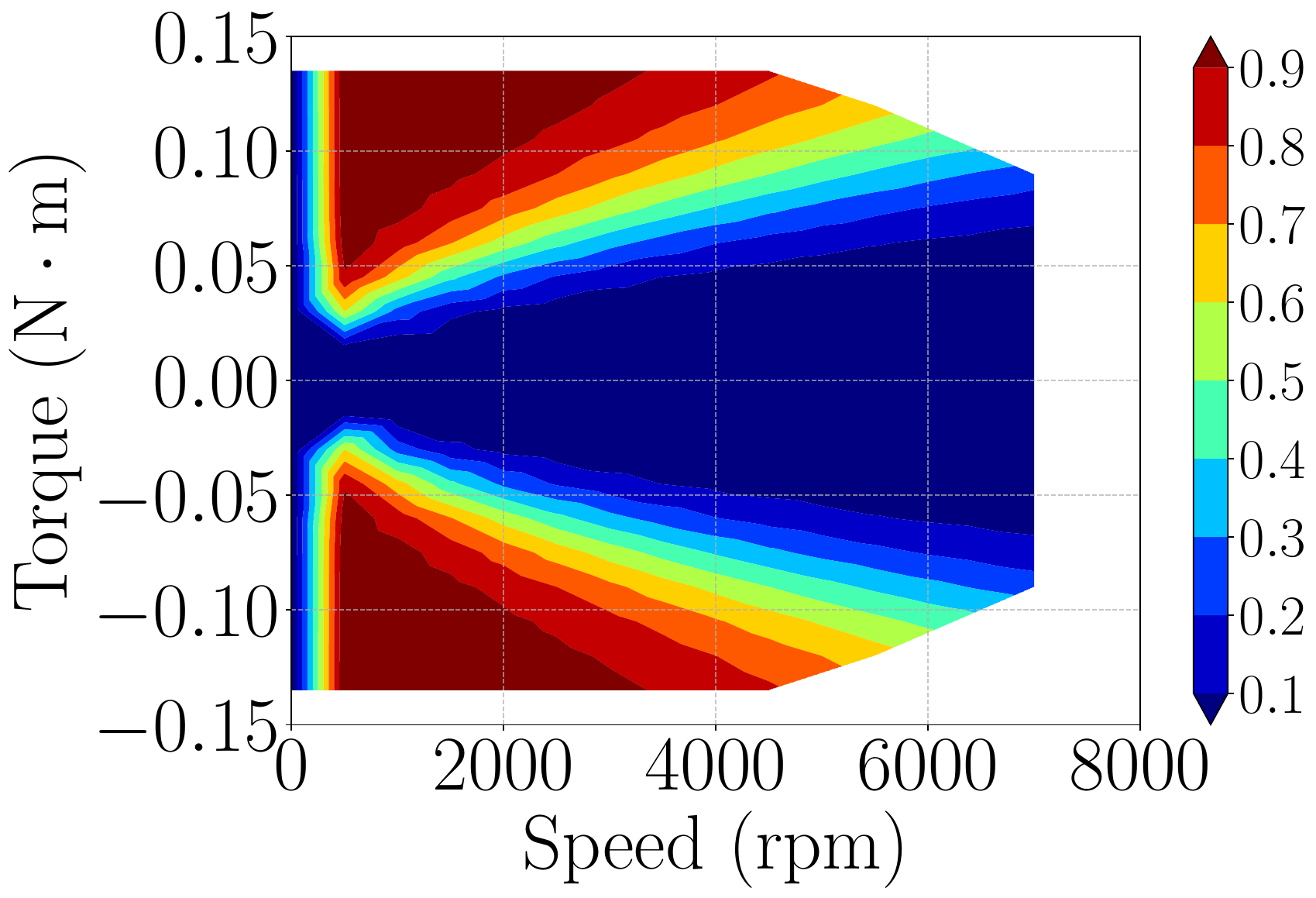}
    \caption{$R_{\mathrm{s}}$.}
    \label{fig:pmsm-grid-sa-Rs}
\end{subfigure}
\\
\begin{subfigure}[b]{0.35\textwidth}
    \centering
    \includegraphics[width=\textwidth]{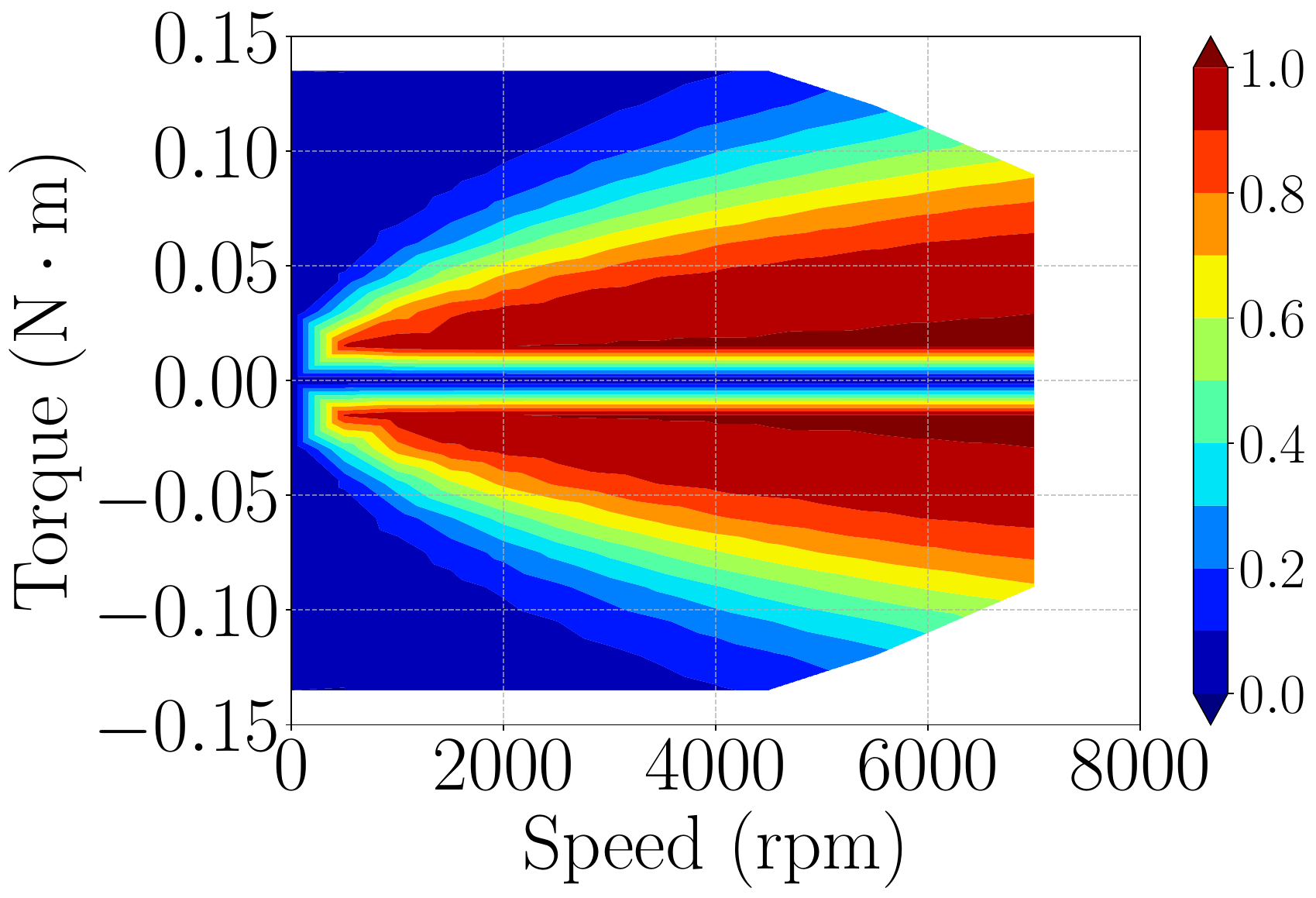}
    \caption{$\lambda$.}
    \label{fig:pmsm-grid-sa-lambda}
\end{subfigure}
\\
\begin{subfigure}[b]{0.35\textwidth}
    \centering
    \includegraphics[width=\textwidth]{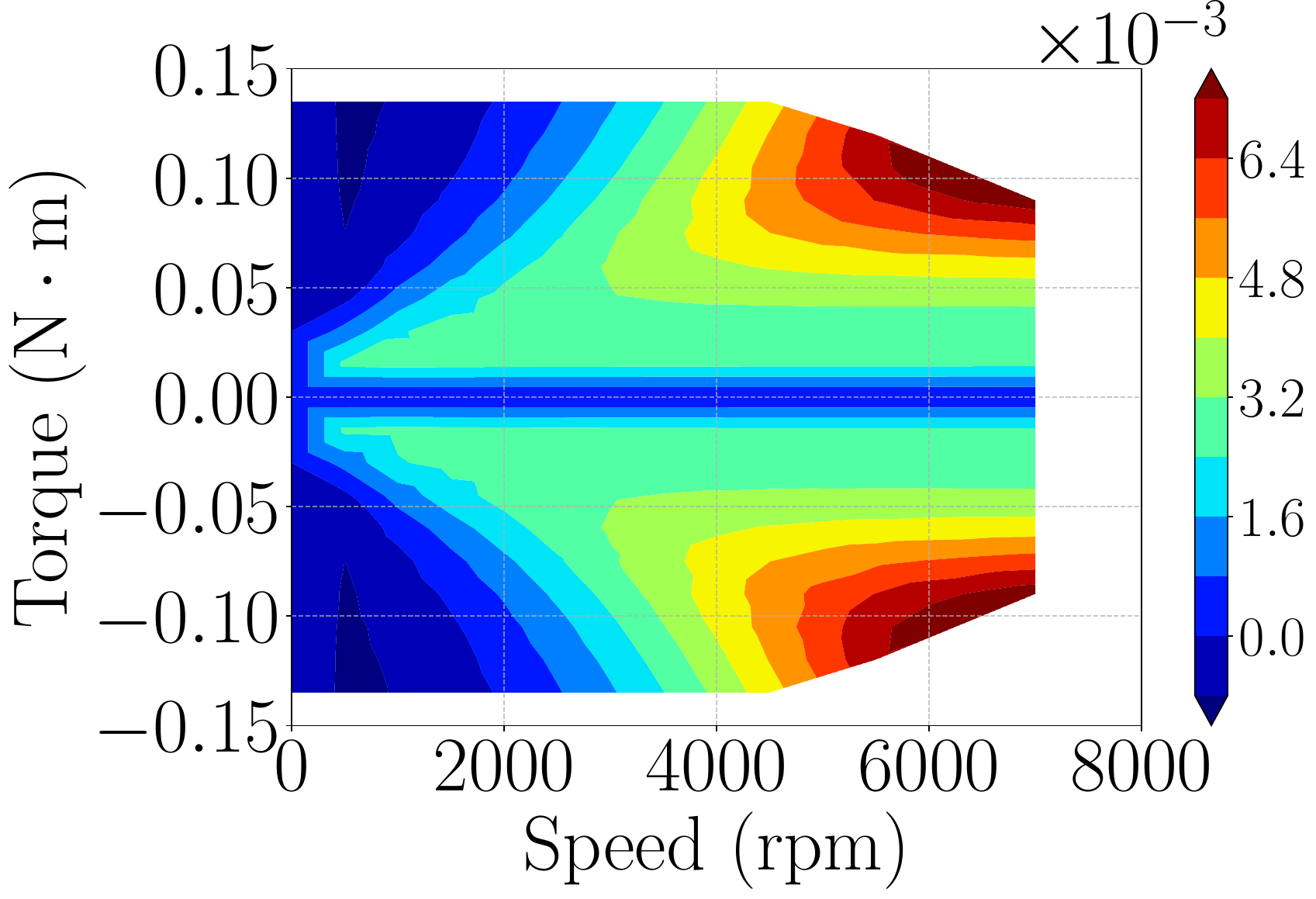}
    \caption{$L_{\text{d}}$.}
    \label{fig:pmsm-grid-sa-Ld}
\end{subfigure}
\\
\begin{subfigure}[b]{0.35\textwidth}
    \centering
    \includegraphics[width=\textwidth]{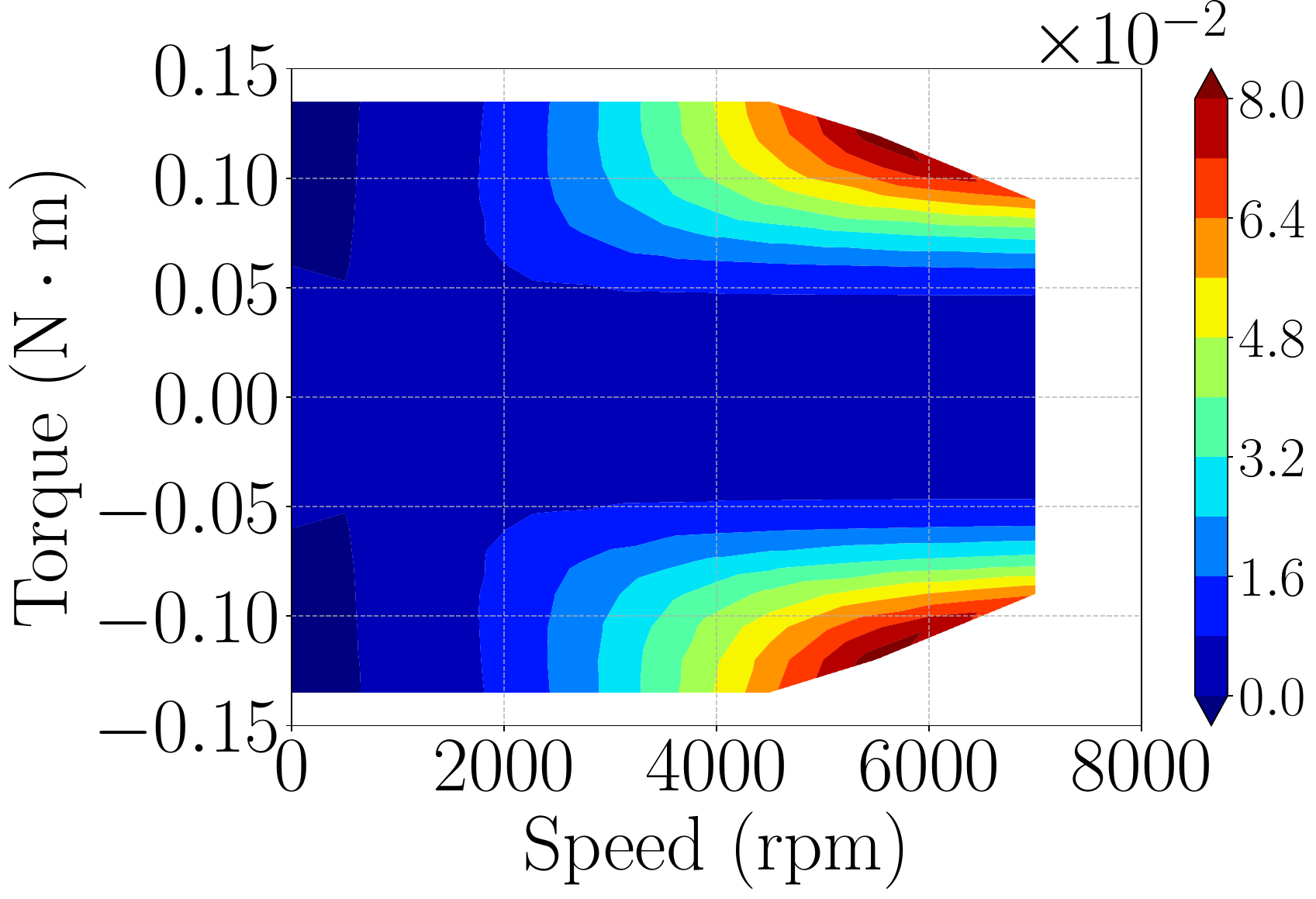}
    \caption{$L_{\text{q}}$.}
    \label{fig:pmsm-grid-sa-Lq}
\end{subfigure}
\caption{\textcolor{black}{Elementwise Sobol' \gls{gsa} of \gls{ecm}'s efficiency map (first-order indices).} }
\label{fig:pmsm-grid-sa}
\end{figure}

\begin{figure}[t!]
    \centering
    \includegraphics[width=0.9\columnwidth]{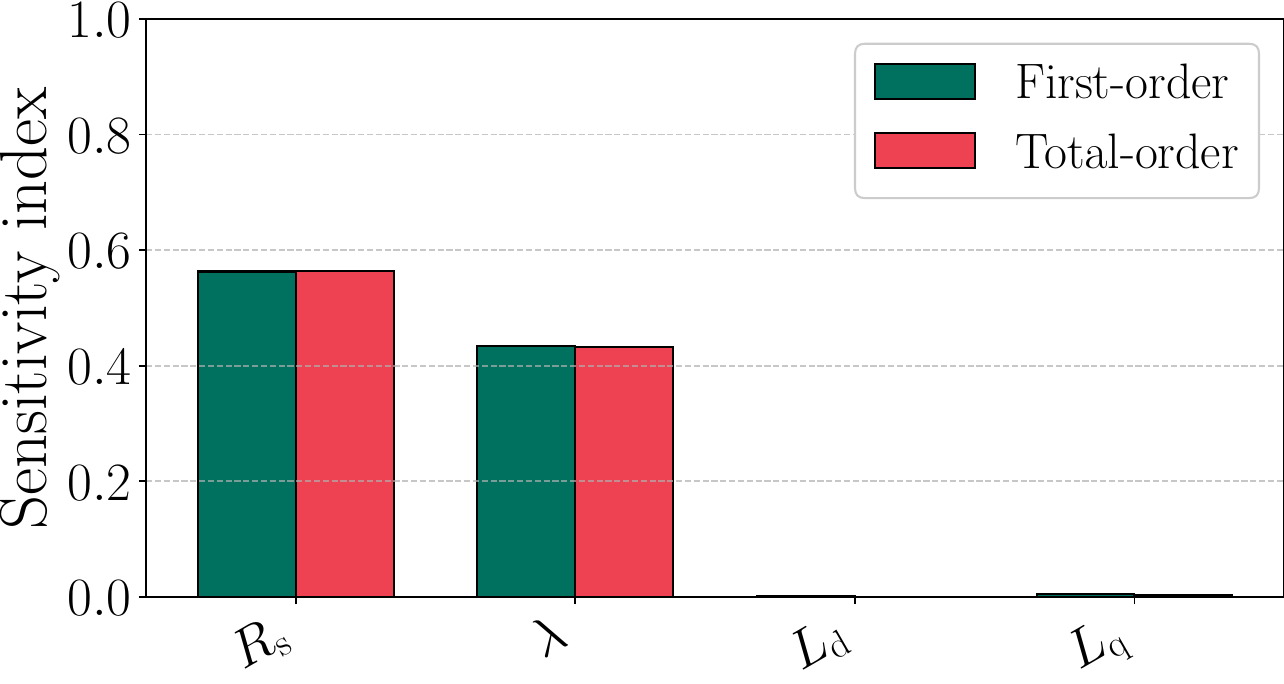}
    \caption{Multivariate \gls{gsa} of \gls{ecm}'s efficiency map.}
    \label{fig:pmsm-grid-mvsa}
\end{figure}

\textcolor{black}{
Figure~\ref{fig:pmsm-grid-sa} illustrates the results of Sobol' \gls{gsa} applied elementwise, which yields one sensitivity map per uncertain parameter. 
Only first-order Sobol' indices are displayed; total-order indices are almost identical, suggesting negligible parameter interactions. 
Figure~\ref{fig:pmsm-grid-mvsa} presents the multivariate \gls{gsa} results. 
The first- and total-order generalized sensitivity indices are nearly identical, further indicating negligible parameter interactions. 
A direct comparison of these figures reveals that the two methods are in agreement that $R_{\text{s}}$ and $\lambda$ have a significant influence on efficiency, while the contribution of $L_{\text{d}}$ is negligible.
However, while in Figure~\ref{fig:pmsm-grid-sa-Lq} $L_{\text{q}}$ appears to have some localized importance for certain operating conditions, its contribution is clearly classified as negligible in Figure~\ref{fig:pmsm-grid-mvsa}.
This discrepancy already highlights an advantage of the multivariate \gls{gsa} approach.
In fact, the Sobol' \gls{gsa} results must be interpreted in the context of the efficiency map’s standard deviation, shown in Figure~\ref{fig:pmsm-grid-uq-std}.
It then becomes obvious that the comparatively high Sobol' index values of $L_{\text{q}}$ in the high-torque and high-speed regions of the efficiency maps are mere numerical artifacts, since the standard deviation in these regions is almost zero.
Hence, these indices do not convey useful sensitivity information.
Multivariate \gls{gsa} altogether avoids this pitfall.
Moreover, comparing the sensitivity maps of Figure~\ref{fig:pmsm-grid-sa} to the generalized indices of Figure~\ref{fig:pmsm-grid-mvsa}, it is obvious that multivariate \gls{gsa} yields results that are much more easily interpretable.
}

To further validate the multivariate \gls{gsa} results, \glspl{mae} in the mean and standard deviation estimates obtained with the full \gls{ecm}, i.e., with random variations in all four parameters, and with a reduced \gls{ecm} with fixed $L_{\text{q}}$ and $L_{\text{d}}$, are computed and presented in Table~\ref{tab:mad-pmsm}. 
The errors in both mean and standard deviation are very low, thus supporting the previous observations. 

\begin{table}[t!]
\centering
\caption{\Glspl{mae} in efficiency map mean and standard deviation estimates between full and reduced \gls{ecm}.}
\label{tab:mad-pmsm}
\begin{tabular}{c c c}
\toprule
Fixed parameters & \gls{mae}, mean & \gls{mae}, st.d. \\
\midrule 
$L_{ \text{d}}$, $L_{\text{q}}$ & $1.08 \cdot 10^{-5}$ & $8.77 \cdot 10^{-6}$ \\ 
\bottomrule 
\end{tabular}
\end{table}

\subsubsection{Sensitivity analysis of efficiency profile}
\label{sec:numexp-pmsm-ecm-drive-cycle}
The \gls{wltp} driving cycle is defined by $N_{\text{op}} = 3601$ operating points $\left(T, \omega_{\text{m}}\right)$, which is one magnitude higher than the efficiency map grid used in Section~\ref{sec:numexp-pmsm-ecm-grid}.
Due to the large number of operating points, \gls{mcs}-based \gls{gsa} becomes too computationally demanding, even for the computationally inexpensive \gls{ecm} evaluations. 
Therefore, only \gls{pce}-based \gls{gsa} results are reported in this section.

\begin{figure}[t!]
\centering
\begin{subfigure}[b]{0.35\textwidth}
    \centering
    \includegraphics[width=\textwidth]{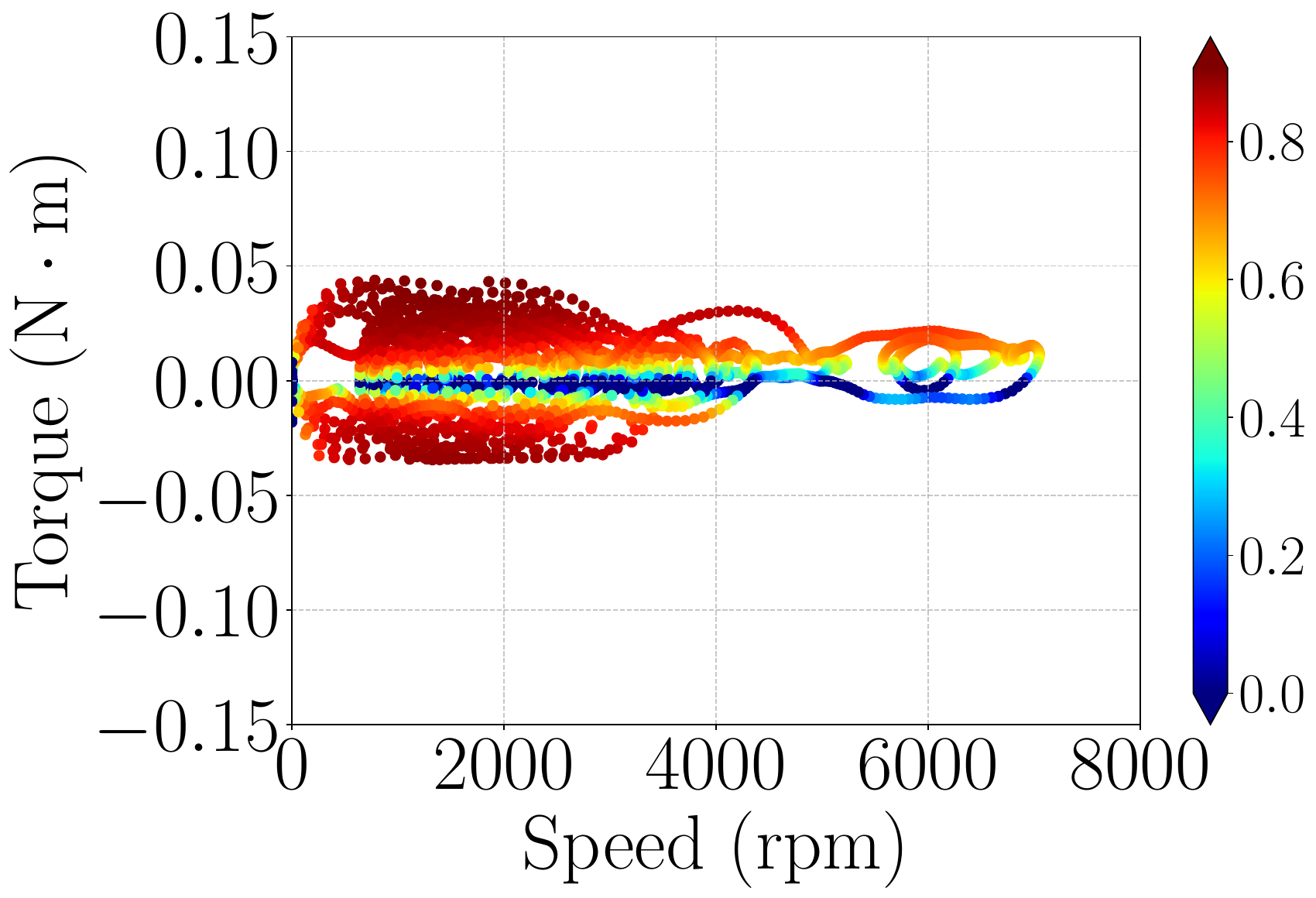}
    \caption{Mean.}
    \label{fig:pmsm_wltp_uq_mean}
\end{subfigure}
\\
\begin{subfigure}[b]{0.35\textwidth}
    \centering
    \includegraphics[width=\textwidth]{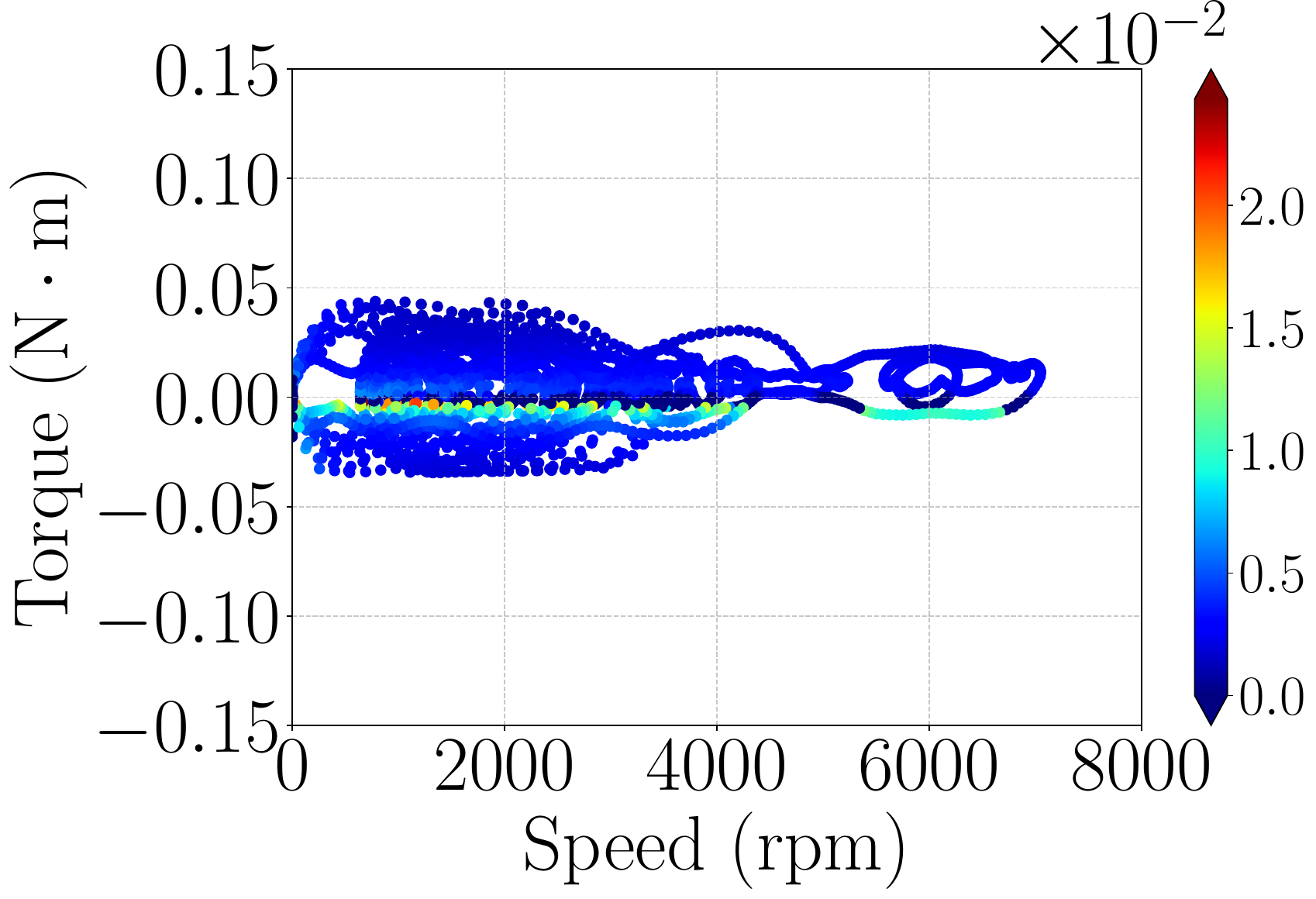}
    \caption{Standard deviation.}
    \label{fig:pmsm_wltp_uq_std}
\end{subfigure}
\caption{\textcolor{black}{Mean and standard deviation of \gls{ecm}'s efficiency profile.}}
\label{fig:pmsm_wltp_uq}
\end{figure}

The mean and standard deviation of the efficiency profile are displayed in Figure~\ref{fig:pmsm_wltp_uq}. 
These results are obtained using \gls{mcs} with sample size $N_{\text{s}}^{\text{MC}} = 2 \cdot 10^5$. 
Almost identical results are obtained using a total-degree \glspl{pce} with maximum polynomial degree $P=4$ and oversampling coefficient $C=3$, which corresponds to a sample size $N_{\text{s}}^{\text{PCE}} = 210$.
The same \gls{pce} is used for the \gls{gsa} results reported in the following.
\textcolor{black}{Similar to Section~\ref{sec:numexp-pmsm-ecm-grid}, the values for $P$ and $C$ are selected over a range of combinations.
Tables~\ref{tab:mean_eff_error_new} and \ref{tab:std_eff_error_new} show the \glspl{mae} in the mean and standard deviation of the efficiency profile for total-degree \glspl{pce} with varying $P$ and $C$.}

\begin{table}[t!]
\centering
\caption{\textcolor{black}{\Glspl{mae} in \gls{ecm}'s mean efficiency profile for \glspl{pce} with varying polynomial degree $P$ and oversampling coefficient $C$.}}
\label{tab:mean_eff_error_new}
\begin{tabular}{ccccc}
\toprule 
$P$ & $C=2$ & $C=3$ & $C=4$ & $C=5$ \\
\midrule 
1 & $2.92\times10^{-6}$ & $1.51\times10^{-6}$ & $1.77\times10^{-6}$ & $2.38\times10^{-6}$ \\
2 & $2.08\times10^{-6}$ & $9.99\times10^{-7}$ & $7.86\times10^{-7}$ & $5.63\times10^{-7}$ \\
3 & $8.26\times10^{-7}$ & $5.98\times10^{-7}$ & $5.86\times10^{-7}$ & $6.08\times10^{-7}$ \\
4 & $6.19\times10^{-7}$ & $5.57\times10^{-7}$ & $6.03\times10^{-7}$ & $5.91\times10^{-7}$ \\
5 & $6.48\times10^{-7}$ & $5.27\times10^{-7}$ & $5.42\times10^{-7}$ & $5.28\times10^{-7}$ \\
\bottomrule
\end{tabular}
\end{table}

\begin{table}[t!]
\centering
\caption{\textcolor{black}{\Glspl{mae} in standard deviation of \gls{ecm}'s efficiency profile for \glspl{pce}  with varying polynomial degree $P$ and oversampling coefficient $C$.}}
\label{tab:std_eff_error_new}
\begin{tabular}{c|cccc}
\toprule 
$P$ & $C=2$ & $C=3$ & $C=4$ & $C=5$ \\
\midrule 
1 & $7.02\times10^{-6}$ & $4.44\times10^{-6}$ & $5.58\times10^{-6}$ & $3.43\times10^{-6}$ \\
2 & $2.44\times10^{-6}$ & $1.75\times10^{-6}$ & $1.61\times10^{-6}$ & $1.61\times10^{-6}$ \\
3 & $1.68\times10^{-6}$ & $1.46\times10^{-6}$ & $1.53\times10^{-6}$ & $1.54\times10^{-6}$ \\
4 & $1.47\times10^{-6}$ & $1.34\times10^{-6}$ & $1.37\times10^{-6}$ & $1.34\times10^{-6}$ \\
5 & $1.41\times10^{-6}$ & $1.40\times10^{-6}$ & $1.37\times10^{-6}$ & $1.34\times10^{-6}$ \\
\bottomrule
\end{tabular}
\end{table}

\begin{figure}[t!]
    \centering
    \begin{subfigure}[b]{0.35\textwidth}
        \centering
        \includegraphics[width=\textwidth]{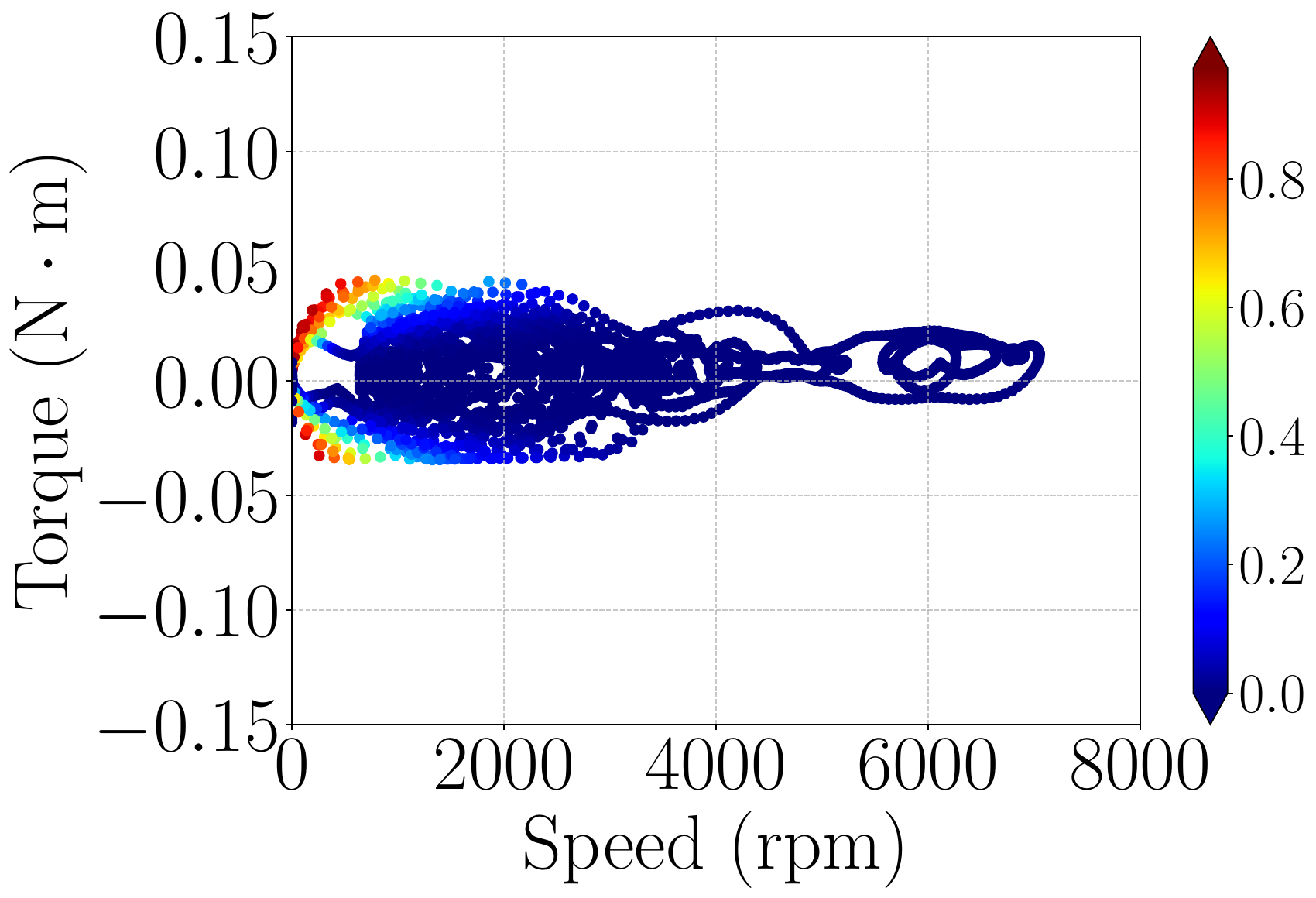}
        \caption{$R_{\mathrm{s}}$.}
    \end{subfigure}
    \hfill
    \begin{subfigure}[b]{0.35\textwidth}
        \centering
        \includegraphics[width=\textwidth]{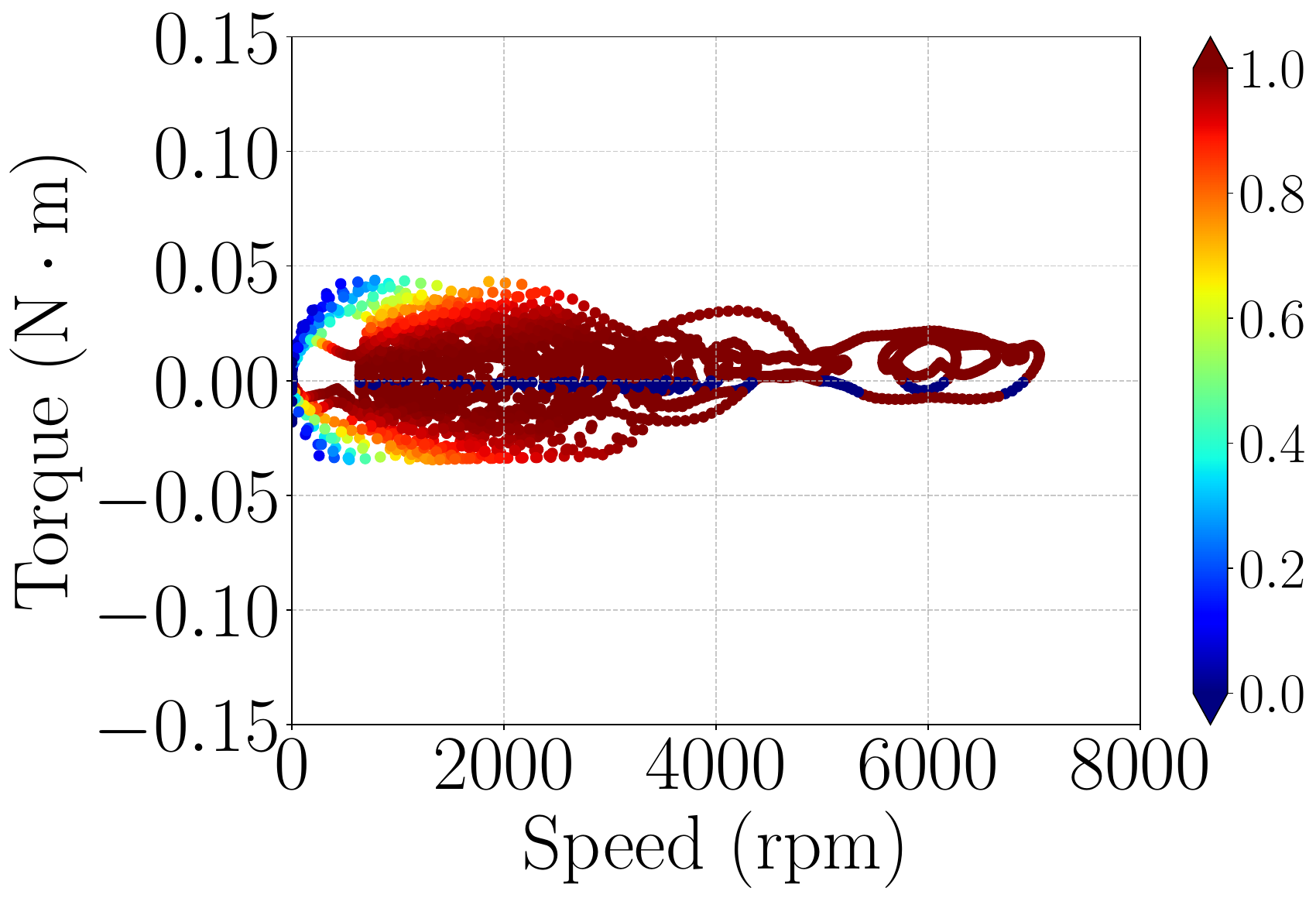}
        \caption{$\lambda$.}
    \end{subfigure}
    \\
    \begin{subfigure}[b]{0.35\textwidth}
        \centering
        \includegraphics[width=\textwidth]{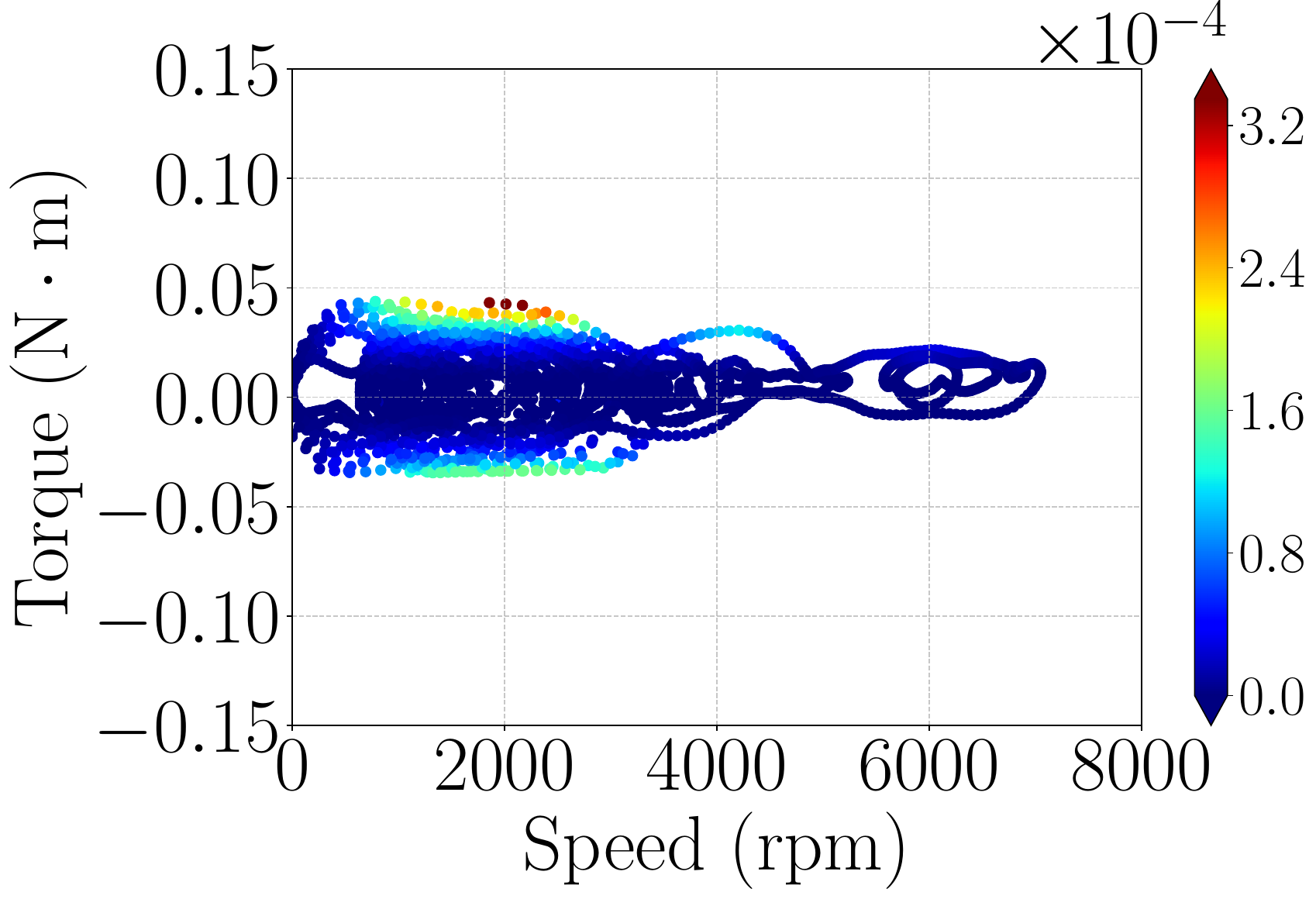}
        \caption{$L_{\text{d}}$.}
    \end{subfigure}
    \hfill
    \begin{subfigure}[b]{0.35\textwidth}
        \centering
        \includegraphics[width=\textwidth]{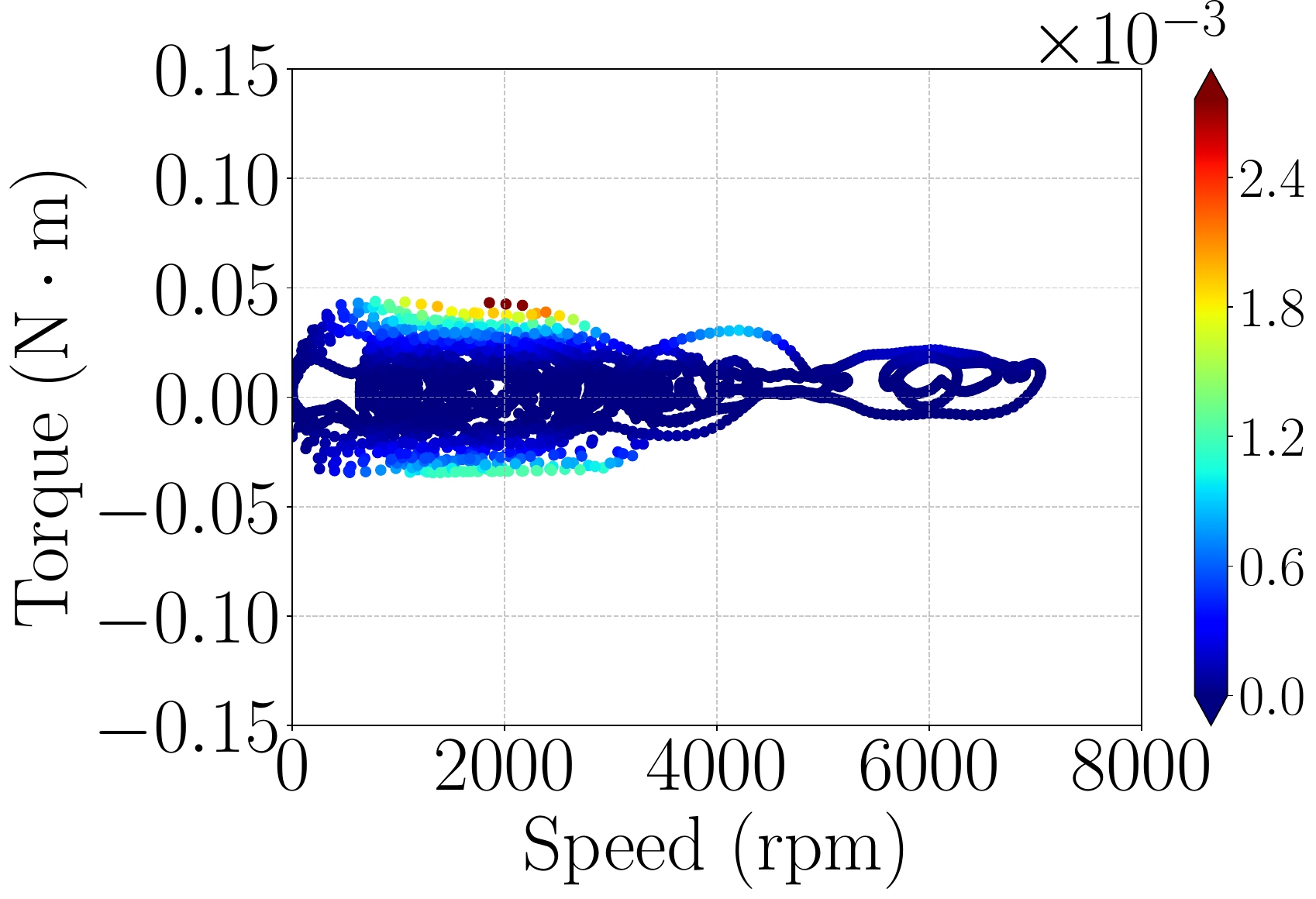}
        \caption{$L_{\text{q}}$.}
    \end{subfigure}
    \caption{\textcolor{black}{Elementwise Sobol' \gls{gsa} of \gls{ecm}'s efficiency profile (first-order indices).}}
    \label{fig:pmsm-wltp-sa}
\end{figure}

\begin{figure}[t!]
\centering
\includegraphics[width=0.9\columnwidth]{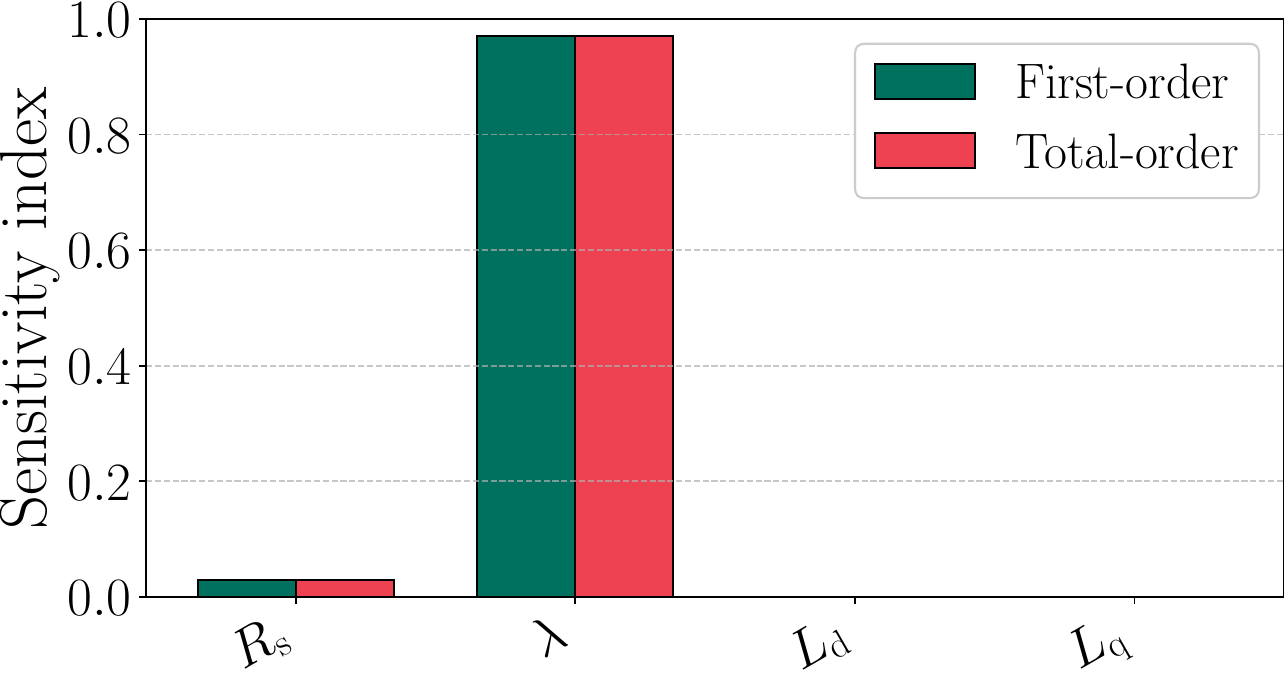}
\caption{Multivariate \gls{gsa} of \gls{ecm}'s efficiency profile.}
\label{fig:pmsm-wltp-mvsa}
\end{figure}

Figure~\ref{fig:pmsm-wltp-sa} presents the elementwise Sobol' \gls{gsa} results for the efficiency profile. 
Figure~\ref{fig:pmsm-wltp-mvsa} shows the corresponding multivariate \gls{gsa} results.
\textcolor{black}{
Both Sobol' and multivariate \gls{gsa} consistently identify  $L_{\text{d}}$ and $L_{\text{q}}$ as non-influential.}
\textcolor{black}{
Despite the agreement of the two methods, multivariate \gls{gsa} presents the relative importance of the uncertain input parameters over the efficiency profile in a much clearer way.
}
\textcolor{black}{
Moreover, $\lambda$ is now the dominant parameter.
These differences compared to Section~\ref{sec:numexp-pmsm-ecm-grid} can be attributed to the fact that the driving cycle does not fully cover the entire operating range. 
In particular, the high-torque regions where $R_{\text{s}}$ and $L_{\text{q}}$ had high sensitivity indices (see Figure~\ref{fig:pmsm-grid-sa}), are not represented in the driving cycle.} 

Table~\ref{tab:mad-pmsm-wltp} reports \glspl{mae} in the mean and standard deviation of the efficiency profile between full and reduced \gls{pmsm} \glspl{ecm}.
The results confirm that parameters deemed as non-influential can be fixed to their nominal values without significantly affecting \gls{uq} accuracy. 

\begin{table}[t!]
\centering
\caption{\glspl{mae} in efficiency profile mean and standard deviation estimates between the full and reduced \gls{ecm}.}
\label{tab:mad-pmsm-wltp}
\begin{tabular}{c c c}
\toprule
Fixed parameters & \gls{mae}, mean & \gls{mae}, st.d. \\
\midrule 
$L_{ \text{d}}$, $L_{\text{q}}$ & $3.0 \cdot 10^{-6}$ & $4.0 \cdot 10^{-6}$ \\ 
\bottomrule 
\end{tabular}
\end{table}

\subsection{Isogeometric analysis model}
\label{sec:numexp-pmsm-iga}
\begin{figure}[t!]
    \centering
    \begin{subfigure}[b]{0.9\columnwidth}
        \centering
        \includegraphics[width=\textwidth]{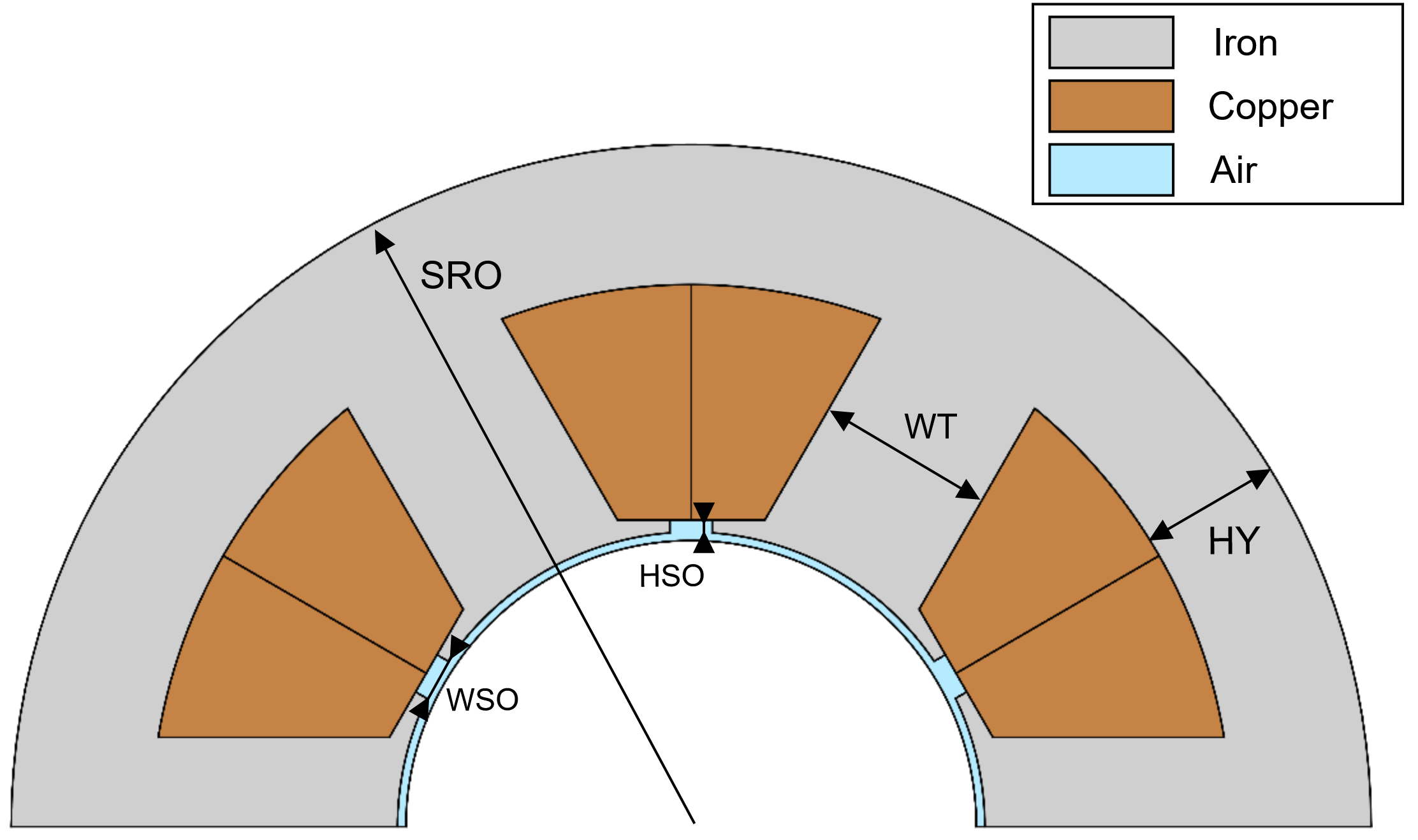}
        \caption{Stator half-geometry.}
    \end{subfigure}
    \begin{subfigure}[b]{0.9\columnwidth}
        \centering
        \includegraphics[width=\textwidth]{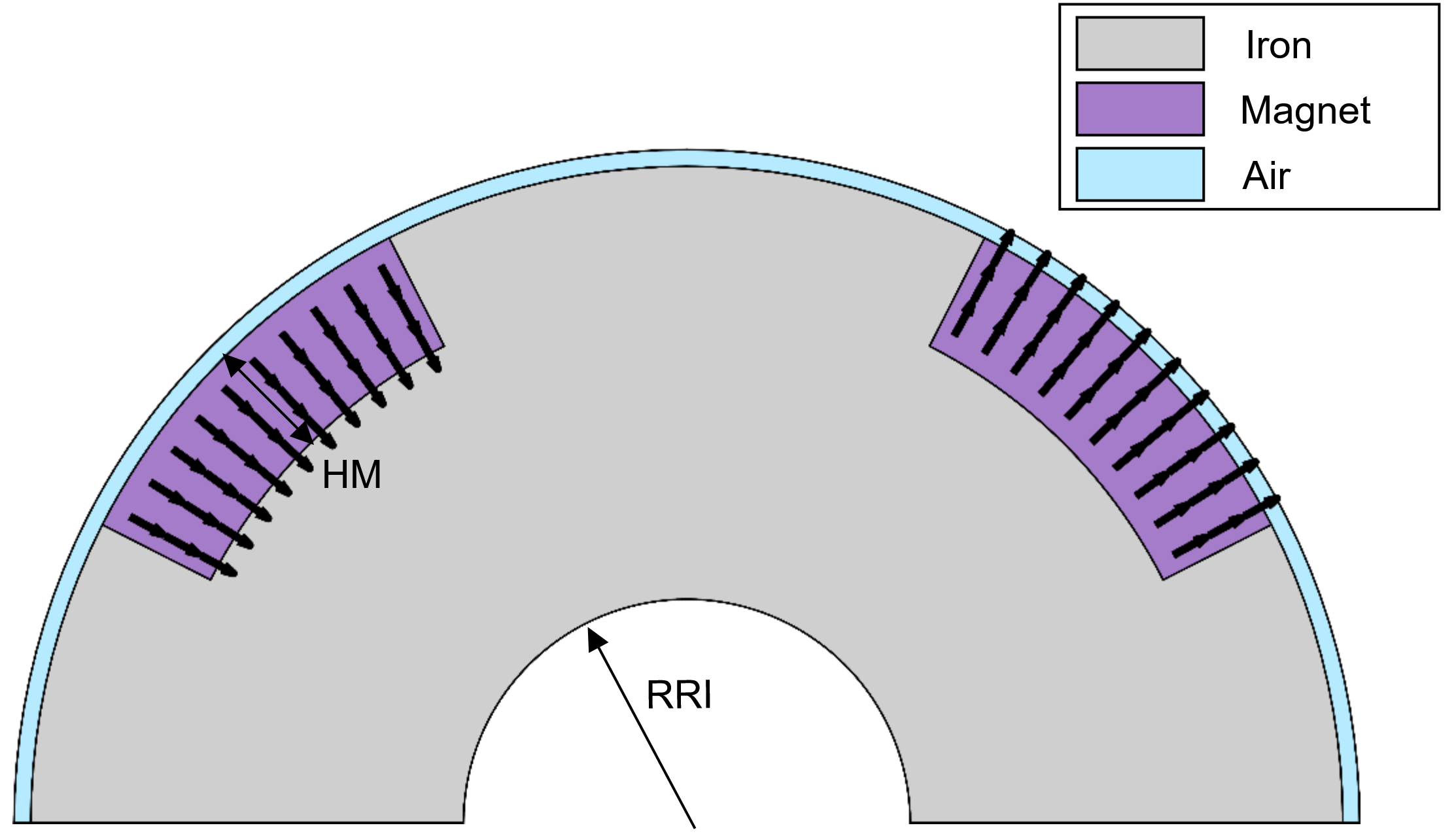}
        \caption{Rotor half-geometry.}
    \end{subfigure}
    \caption{\gls{pmsm} half-geometry in two dimensions.}
    \label{fig:geometry}
\end{figure}

The \gls{iga} model of the \gls{pmsm} is implemented in \texttt{MATLAB} using the \texttt{GeoPDEs} package \cite{vazquez2016new}. 
The model is based on standard finite element modeling \cite{salon1995finite}, where the basis functions coincide with the \gls{cad} spline basis~\cite{wiesheu2025combined}. 
Eleven model parameters are considered to be uncertain, each assumed to vary uniformly within $\pm 5\%$ of its nominal value. 
The parameters and their nominal values are listed in Table~\ref{tab:num-model-parameters}. 
The geometry of the \gls{pmsm} is illustrated in Figure~\ref{fig:geometry}, along with the uncertain geometric parameters, with the exception of the magnet circumference ratio (MR). 

\begin{table}[t!]
\centering
\caption{Parameters of \gls{pmsm} \gls{iga} model.}
\label{tab:num-model-parameters}
\resizebox{\columnwidth}{!}{
\begin{tabular}{lccc}
\toprule
Parameter & Symbol & Nominal Value & Units\\
\midrule
Stator outer radius & SRO & $0.0565$ & $\mathrm{m}$ \\
Yoke height & HY & $0.00116$ & $\mathrm{m}$ \\ 
Slot opening height & HSO & $0.00098$ & $\mathrm{m}$ \\ 
Tooth width & WT & $0.00148$ & $\mathrm{m}$ \\ 
Slot opening width & WSO & $0.00323$ & $\mathrm{m}$ \\ 
Magnet height & HM & $0.00435$ & $\mathrm{m}$ \\
Rotor inner radius & RRI & $0.008$ & $\mathrm{m}$ \\ 
Magnet circumference ratio & MR & $0.5$ & -- \\
$BH$-curve scaling factor & SF & $1.0$  & -- \\
Phase resistance & $R_\text{s}$ & $4.475$ & $\Omega$ \\
Magnet remanence & $B_{\text{r}}$ & $0.41$ & $\mathrm{T}$ \\
\bottomrule
\end{tabular}
}
\end{table}

Compared to the \gls{ecm} used previously, the \gls{iga} model introduces a number of challenges. 
First, due to its much higher computational cost, \gls{mcs} is rendered unusable, both for statistics estimates and for \gls{gsa}.
Second, it becomes too computationally expensive to evaluate the model over its full operating range, as well as for the \gls{wltp} driving cycle.
Last, the comparatively high input dimensionality ($N=11$) results in large numbers of total-degree \gls{pce} terms, even for moderate polynomial degrees $P$. 
In turn, generating a sufficiently large sample amounts to an undesirable computational cost.
To render the computational cost tractable, the \gls{pmsm} is operated solely in motoring mode and \gls{gsa} is confined to efficiency maps only. 
Using a grid of $N_{\text{op}} = 116$ operating points $(T, \omega_{\text{m}})$, a single efficiency map is computed in approximately $800$ seconds.
Additionally, the use of \gls{pce} is enabled by adaptive basis construction algorithms \cite{loukrezis2020robust, loukrezis2025multivariate}.
Samples of increasing sizes are employed, up to a maximum size $N_{\text{s}}^{\max} = 3300$.

\begin{figure}[t!]
\centering
\begin{subfigure}[b]{0.35\textwidth}
    \centering
    \includegraphics[width=\textwidth]{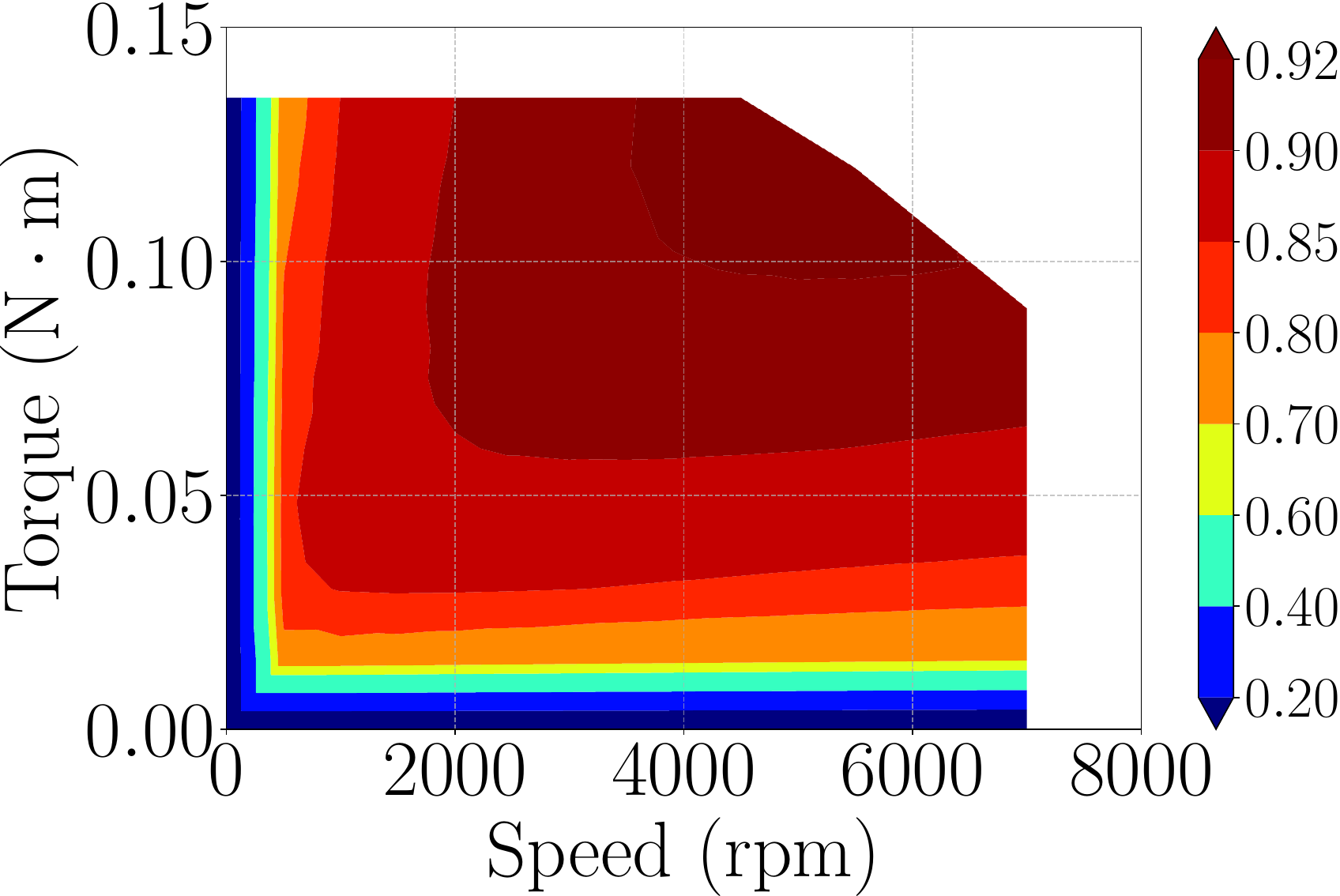}
    \caption{Mean.}
    \label{fig:pmsm-n-grid-uq-mean}
\end{subfigure}
\
\begin{subfigure}[b]{0.35\textwidth}
    \centering
    \includegraphics[width=\textwidth]{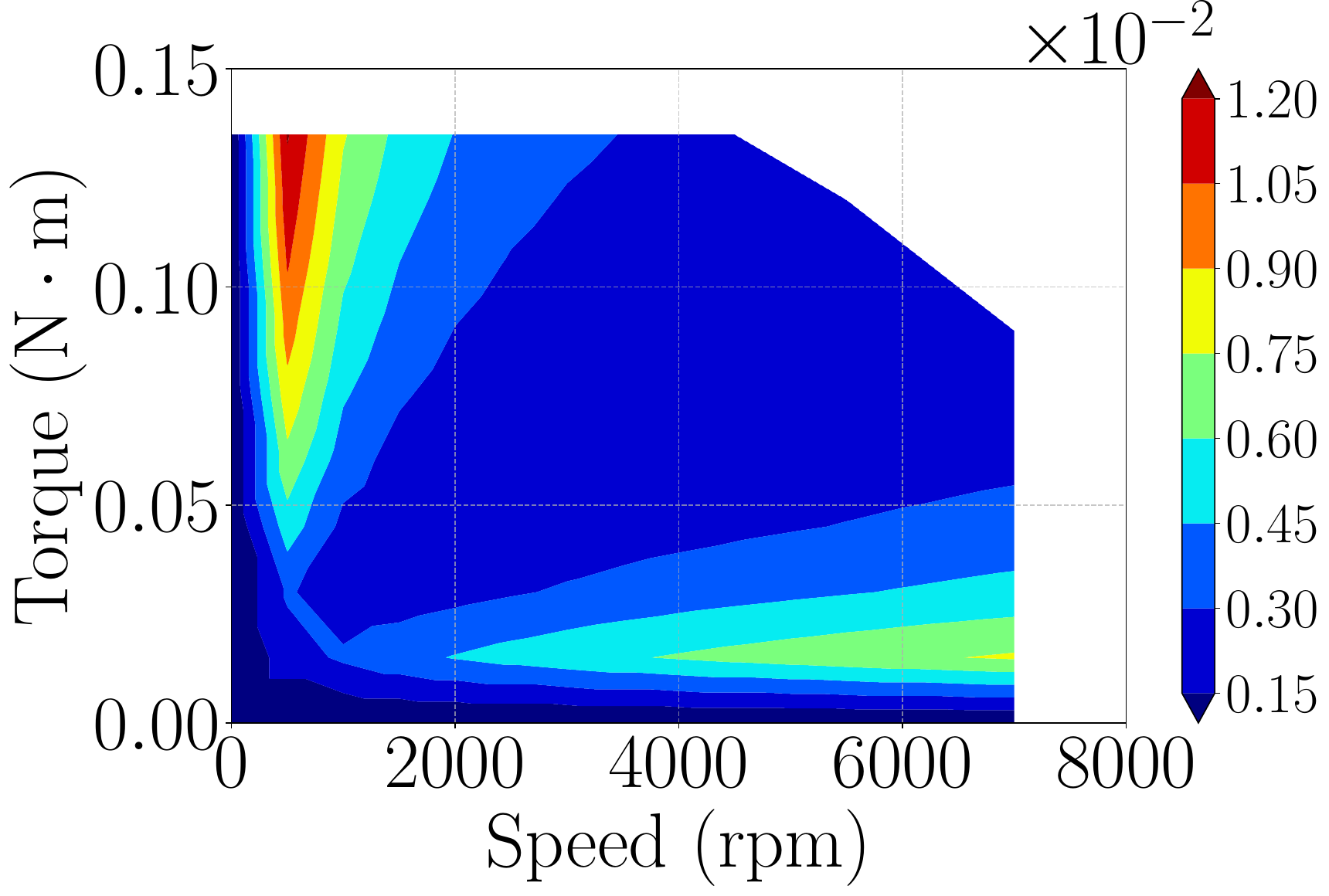}
    \caption{Standard deviation.}
    \label{fig:pmsm-n-grid-uq-std}
\end{subfigure}
\hfill 
\begin{subfigure}[b]{0.35\textwidth}
\includegraphics[width=\textwidth]{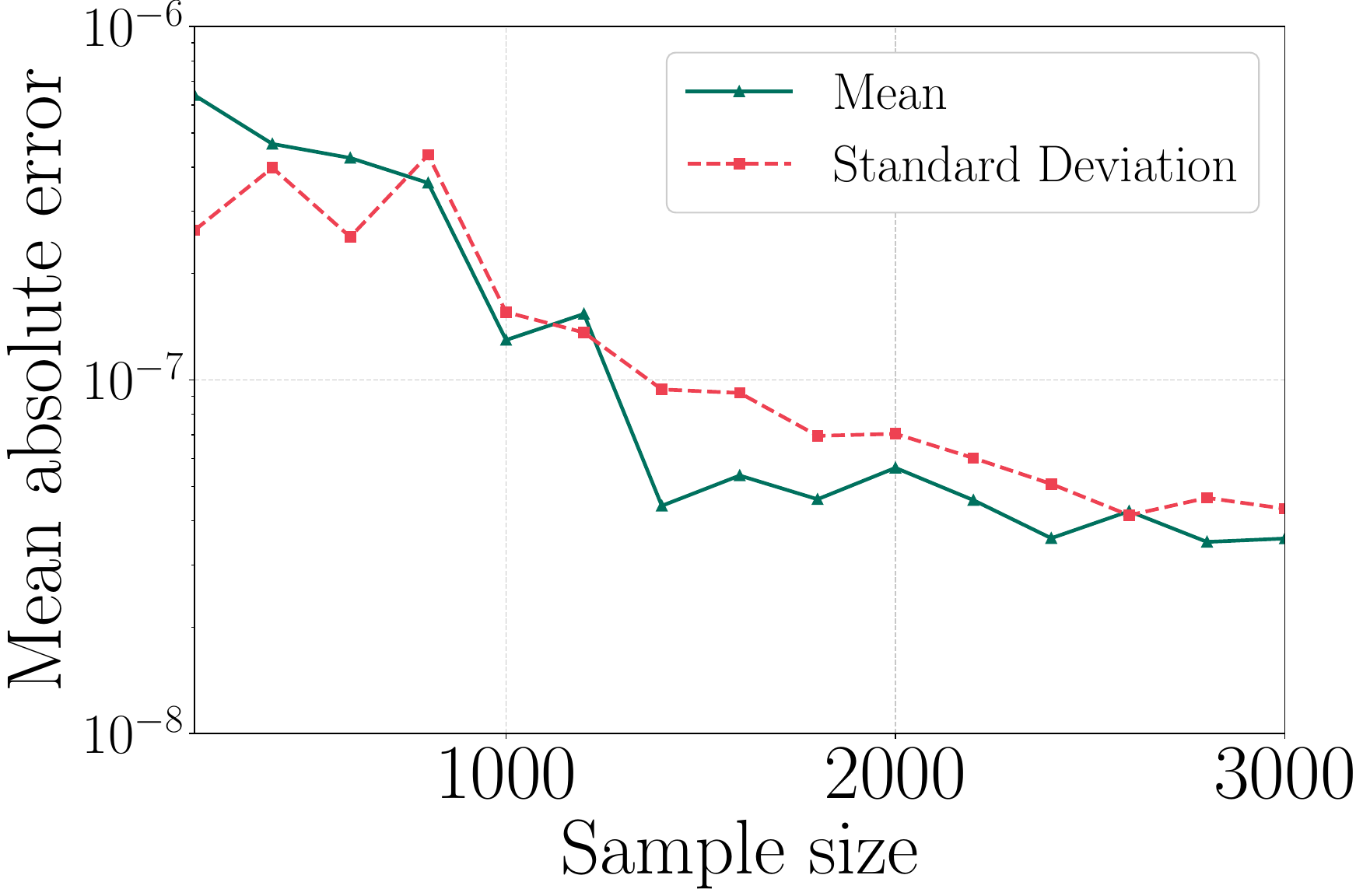}
\caption{Statistics errors.}
\label{fig:pmsm-n-error}
\end{subfigure}
\caption{Mean and standard deviation of \gls{iga} model's efficiency map, with \glspl{mae} for increasing sample sizes. The \glspl{mae} are computed with respect to reference values obtained with a \gls{pce} trained on the dataset of maximum size $N_{\text{s}, \max}=3300$.}
\label{fig:pmsm-n-grid-uq}
\end{figure}

\begin{figure*}[t!]
    \centering
    \begin{subfigure}[b]{0.32\textwidth}
        \centering
        \includegraphics[width=\textwidth]{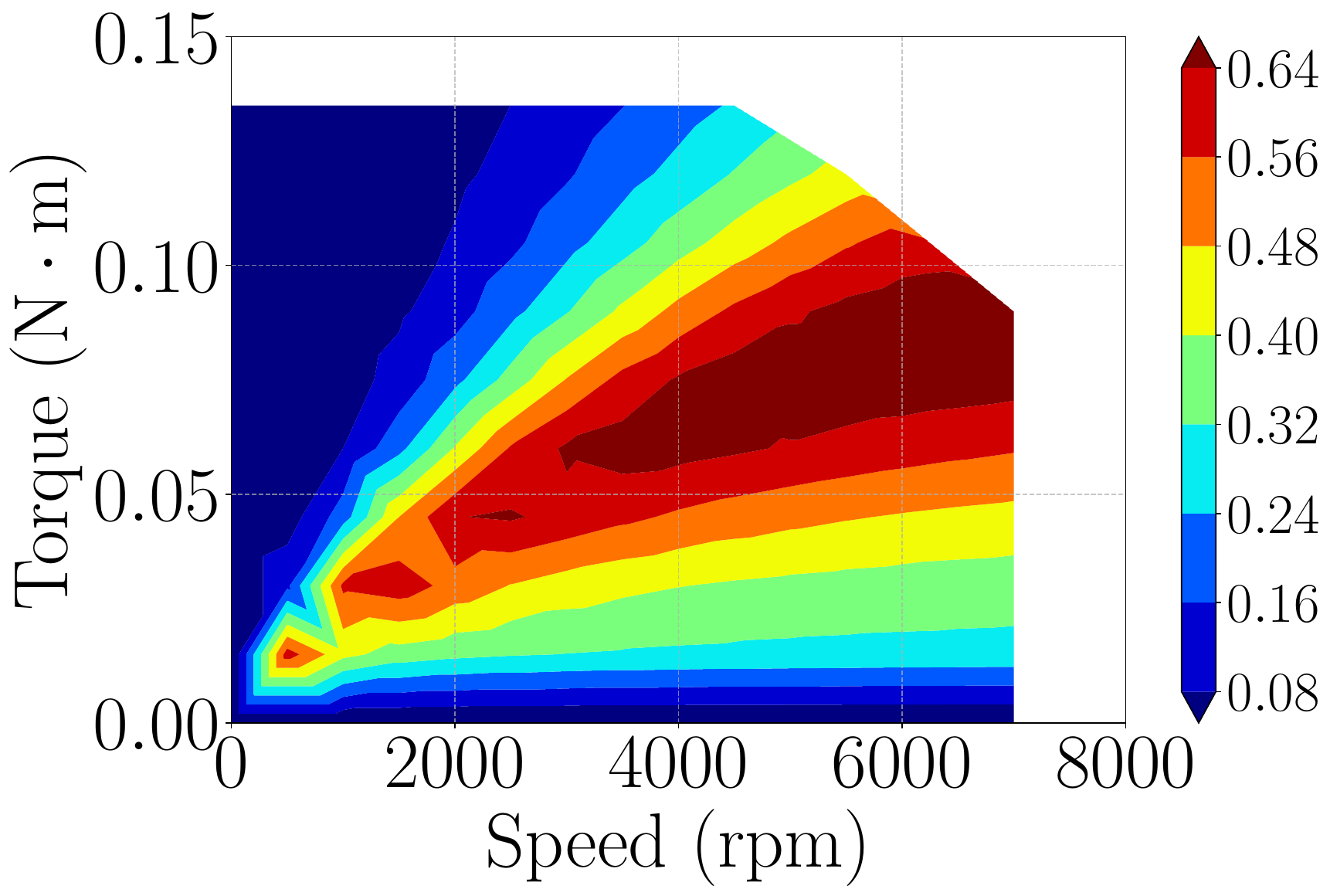}
        \caption{SRO.}
    \end{subfigure}
    \hfill 
    \begin{subfigure}[b]{0.32\textwidth}
        \centering
        \includegraphics[width=\textwidth]{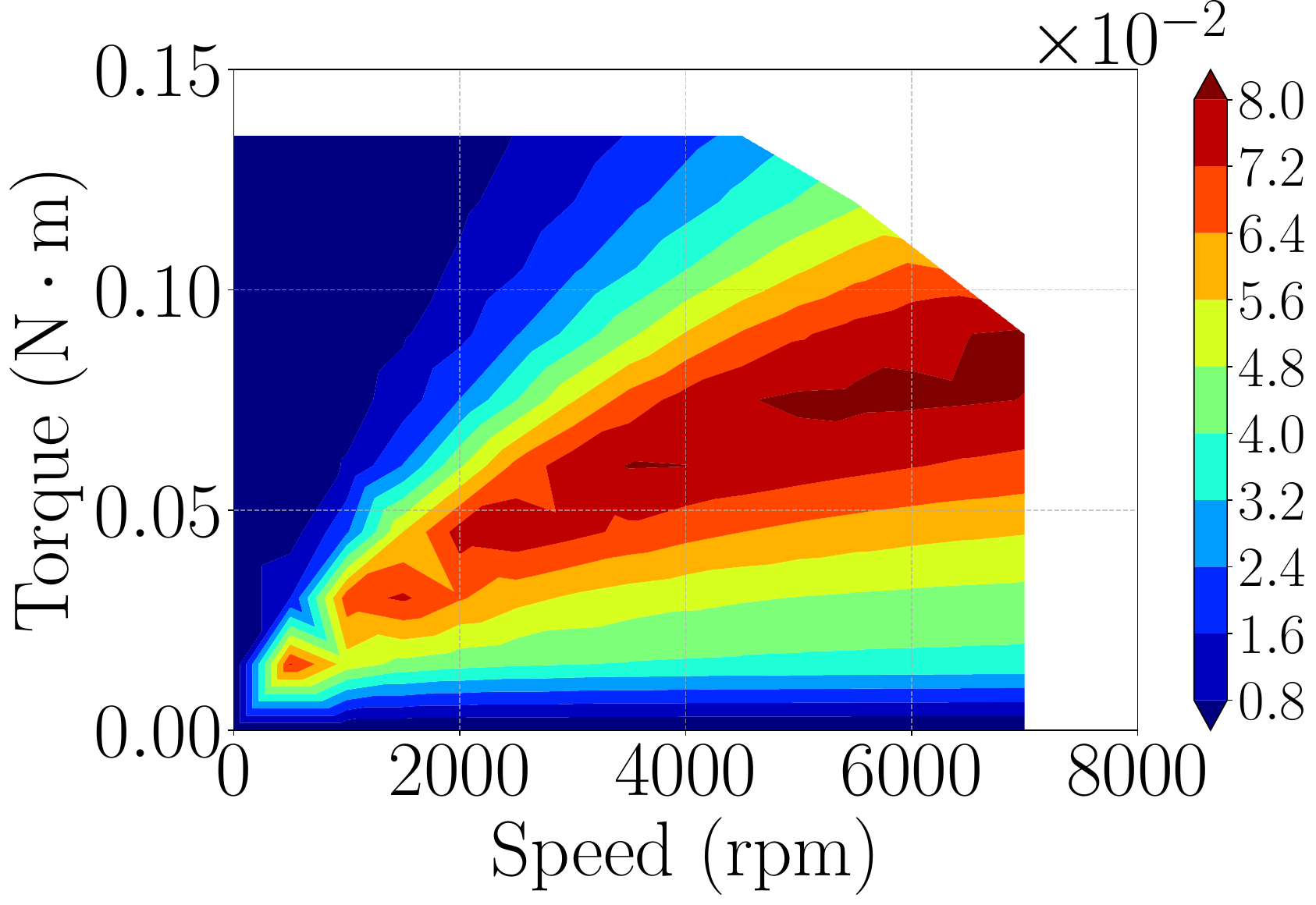}
        \caption{HY.}
    \end{subfigure}
    \hfill
    \begin{subfigure}[b]{0.32\textwidth}
        \centering
        \includegraphics[width=\textwidth]{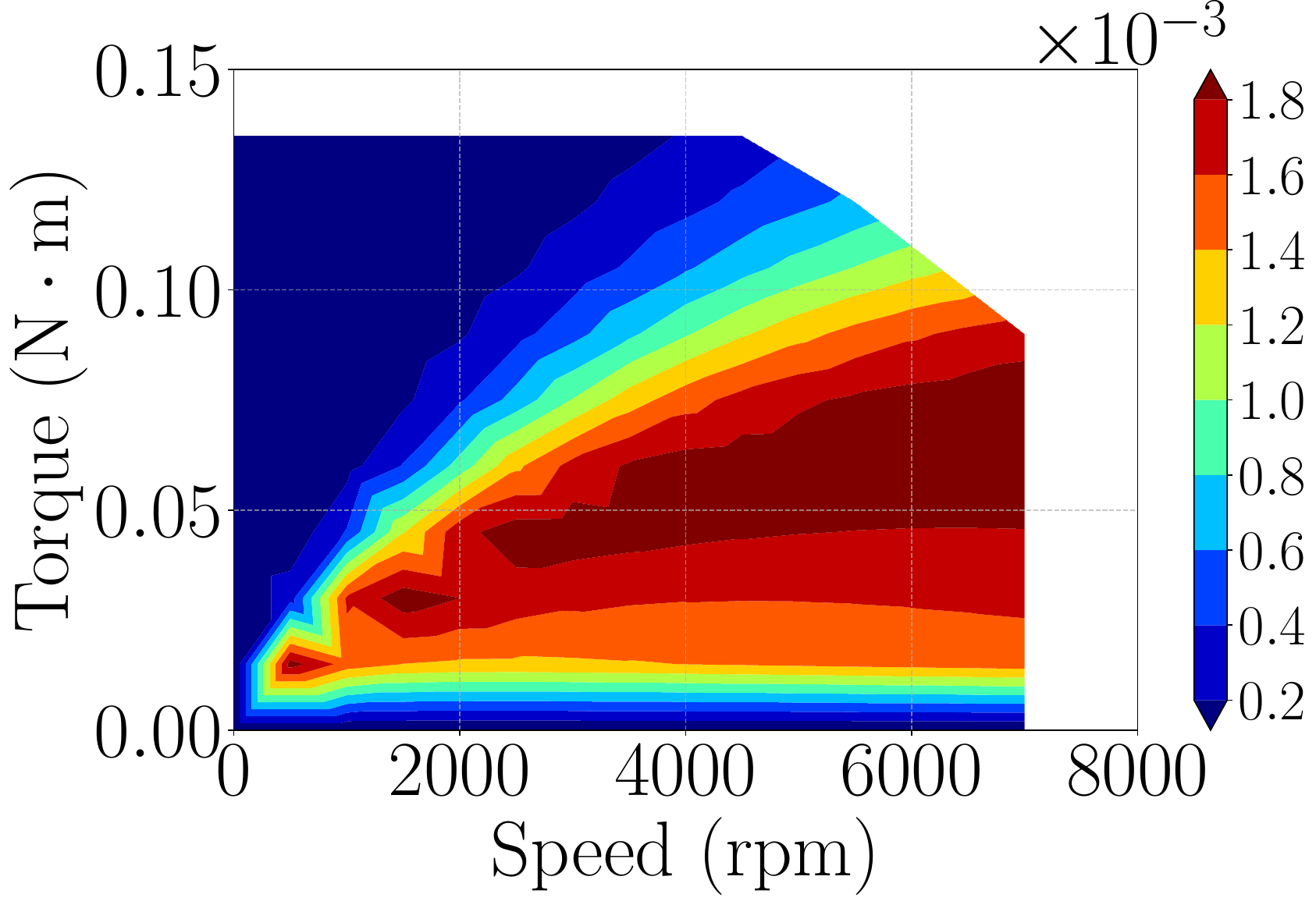}
        \caption{HSO.}
    \end{subfigure}
    \\
    \begin{subfigure}[b]{0.32\textwidth}
        \centering
        \includegraphics[width=\textwidth]{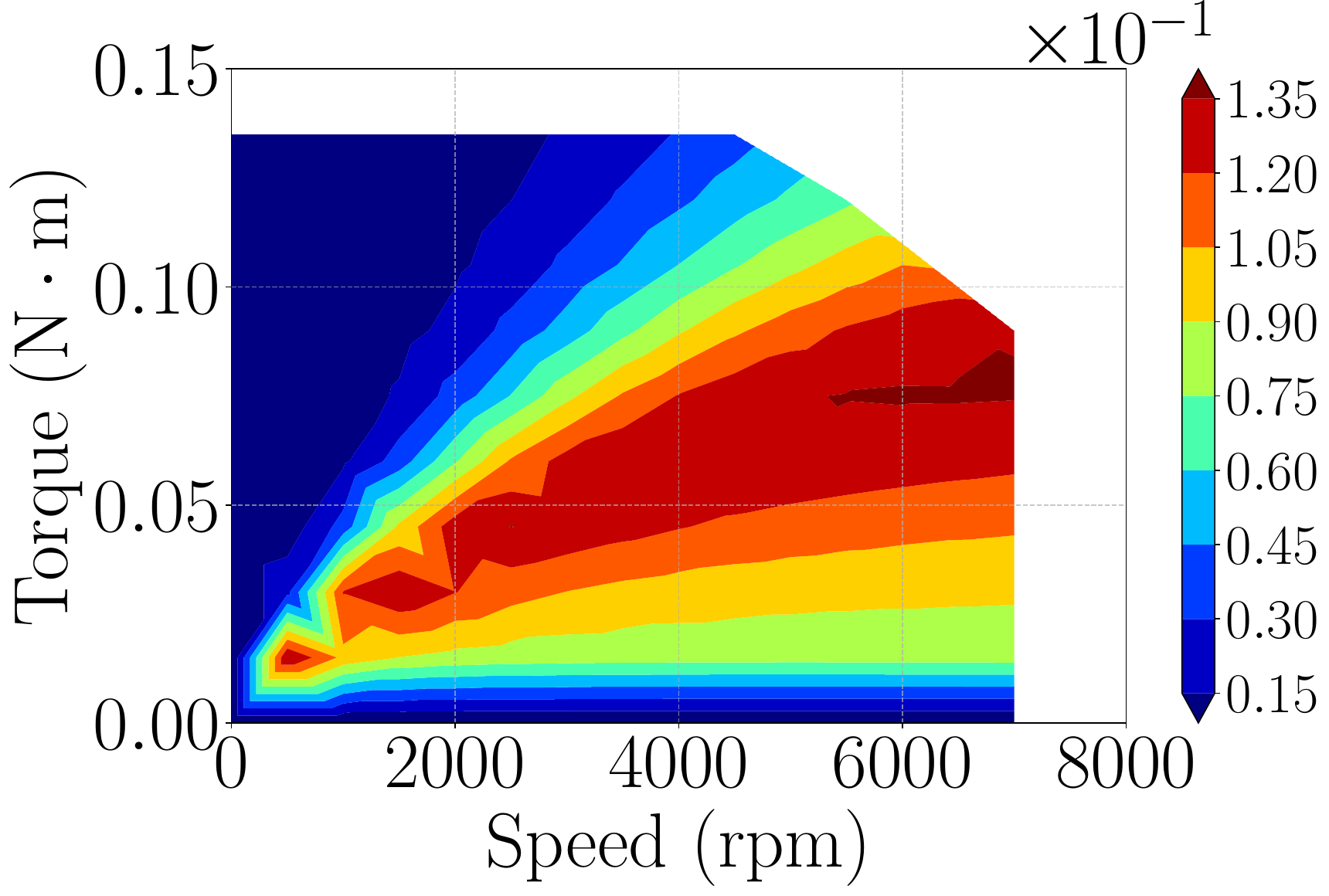}
        \caption{WT.}
    \end{subfigure}
    \hfill 
    \begin{subfigure}[b]{0.32\textwidth}
        \centering
        \includegraphics[width=\textwidth]{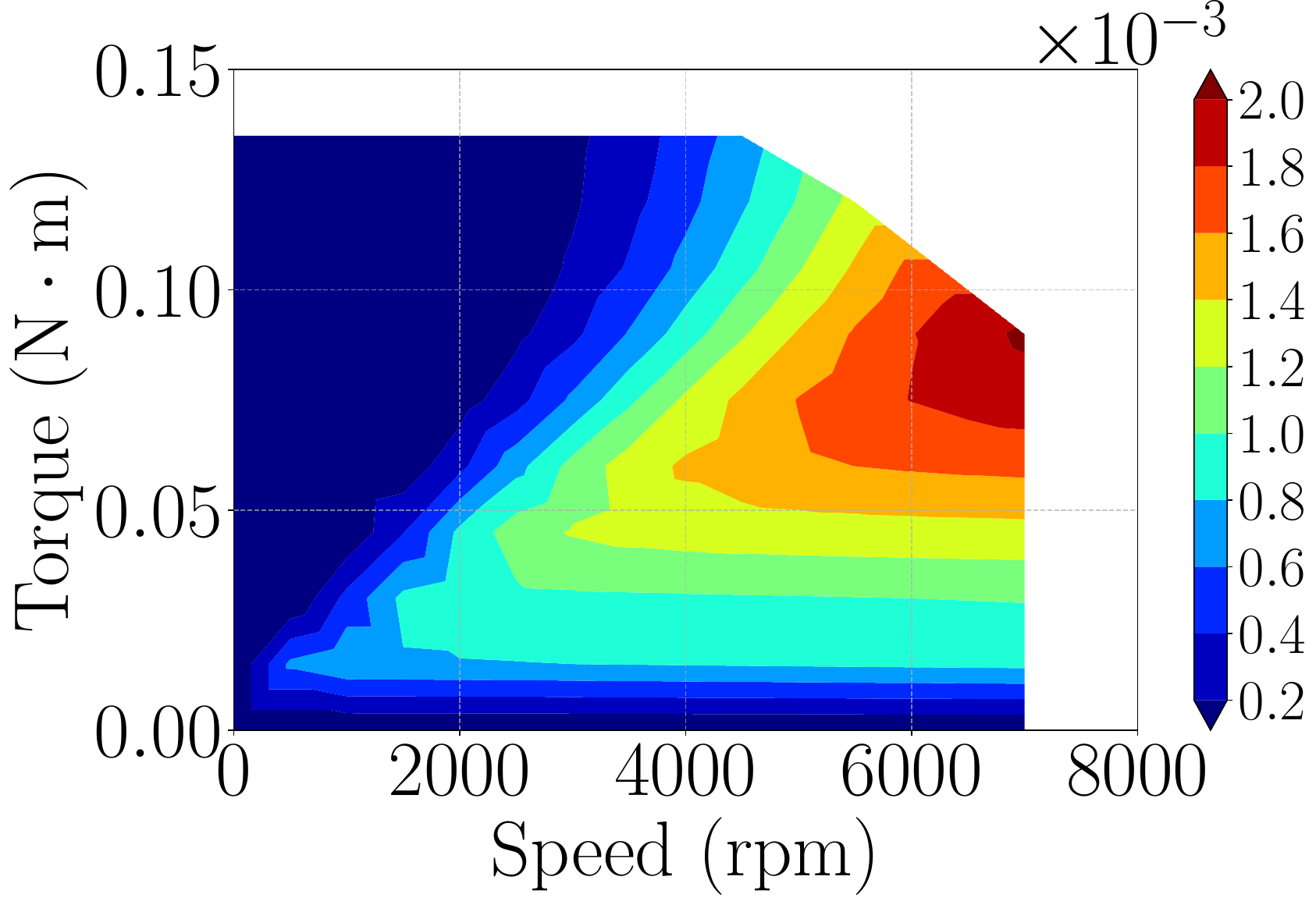}
        \caption{WSO.}
    \end{subfigure}
    \hfill
    \begin{subfigure}[b]{0.32\textwidth}
        \centering
        \includegraphics[width=\textwidth]{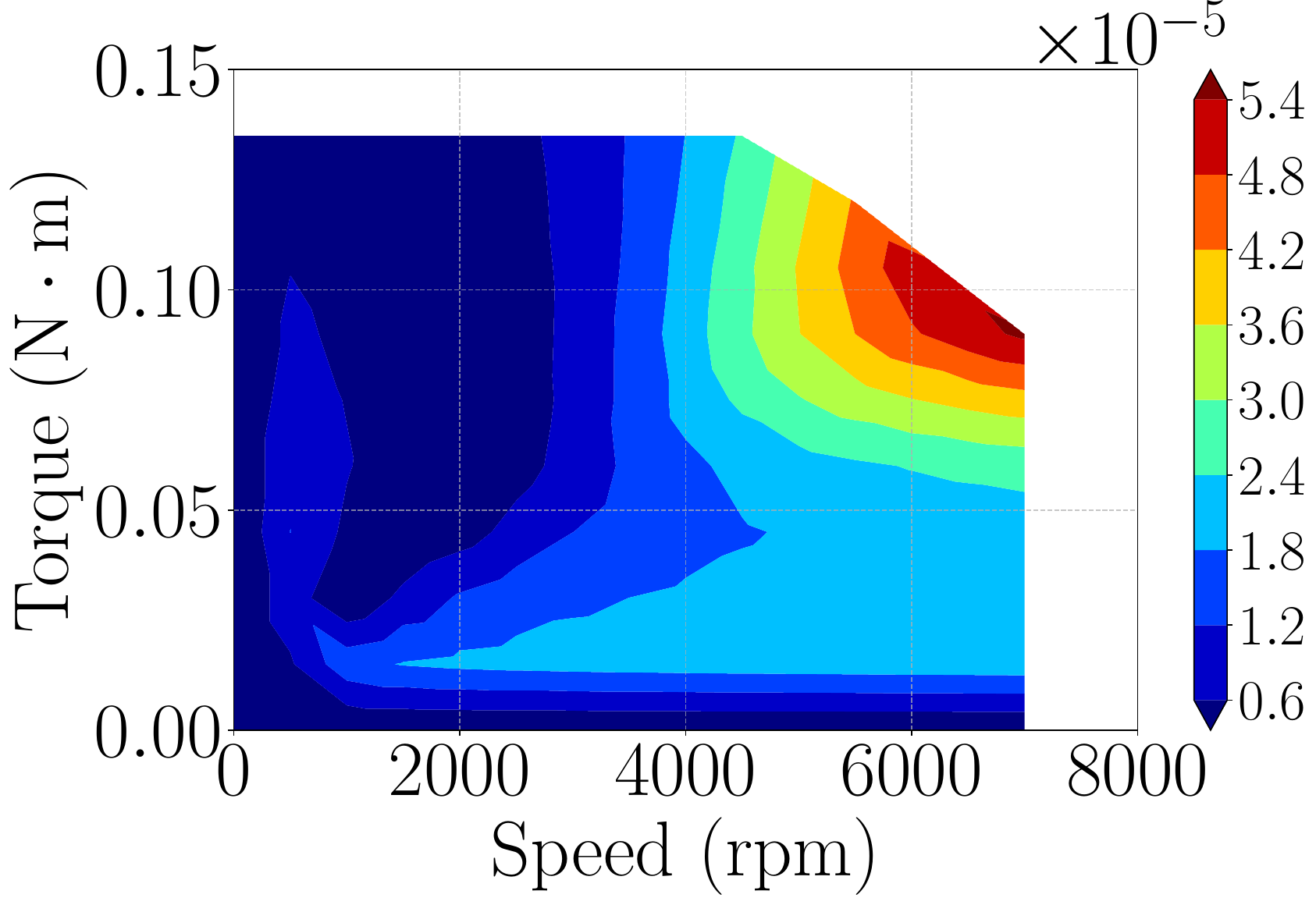}
        \caption{HM.}
    \end{subfigure}
    \\ 
    \begin{subfigure}[b]{0.32\textwidth}
        \centering
        \includegraphics[width=\textwidth]{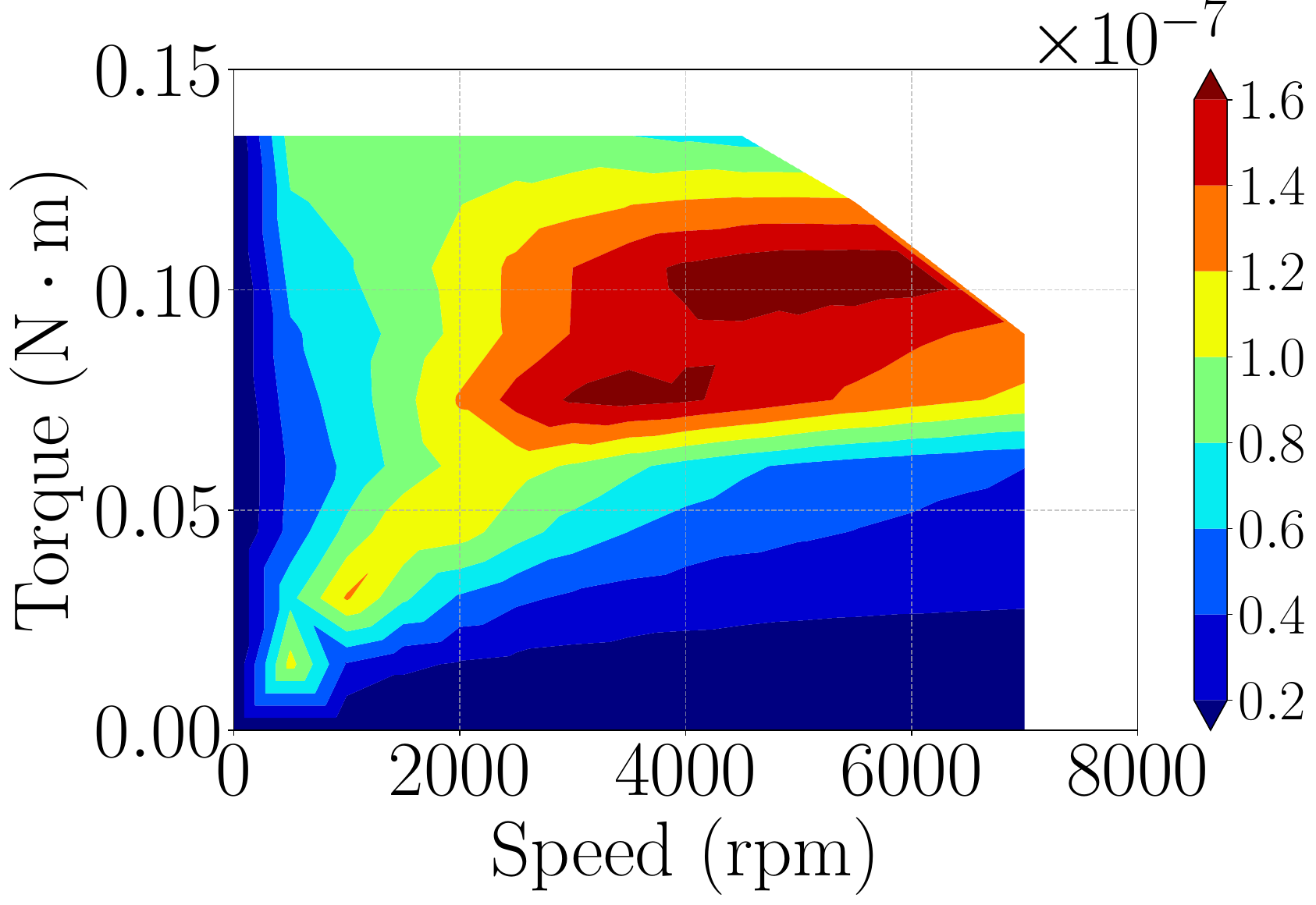}
        \caption{RRI.}
    \end{subfigure}
    \hfill
    \begin{subfigure}[b]{0.32\textwidth}
        \centering
        \includegraphics[width=\textwidth]{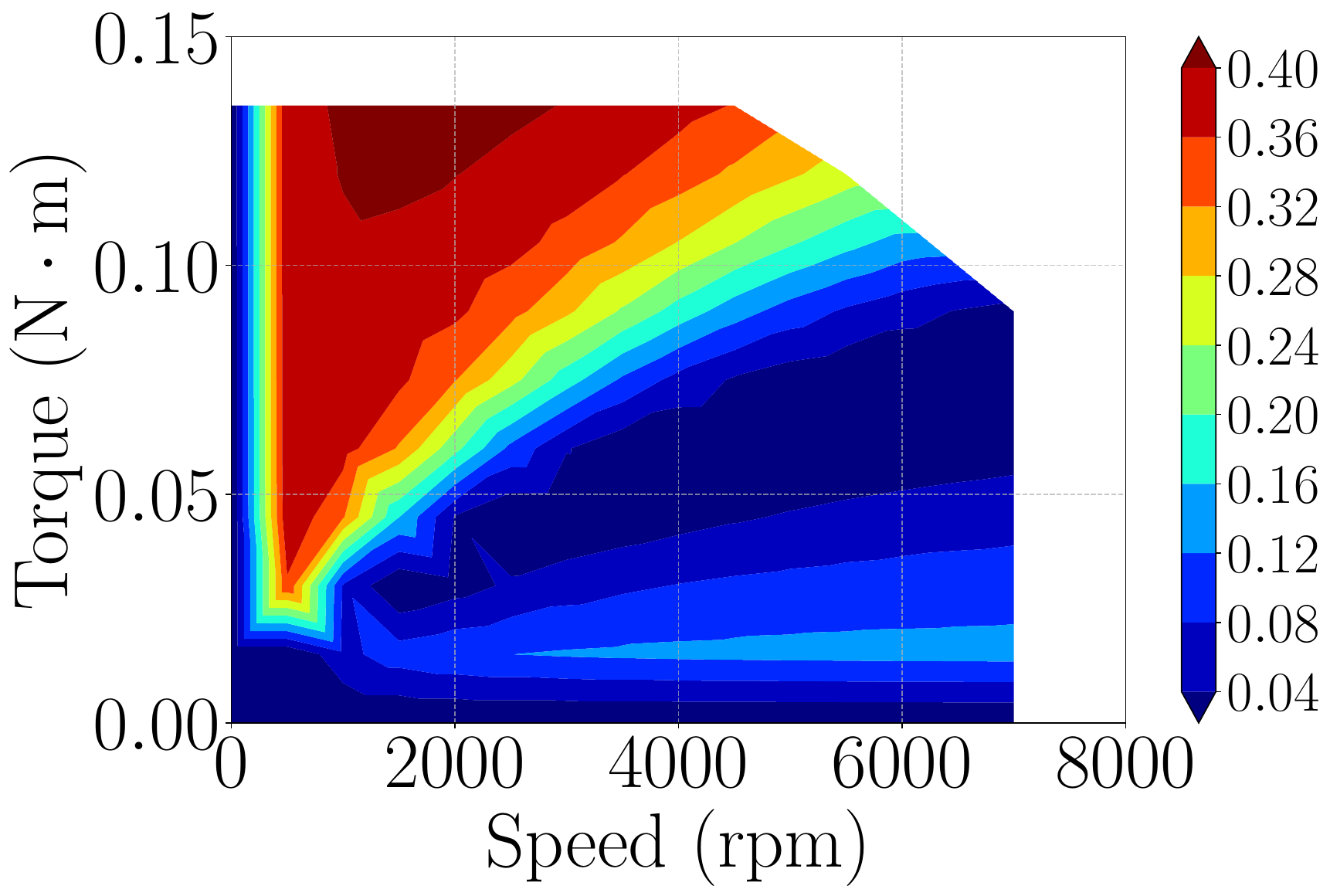}
        \caption{MR.}
    \end{subfigure}
    \hfill
    \begin{subfigure}[b]{0.32\textwidth}
        \centering
        \includegraphics[width=\textwidth]{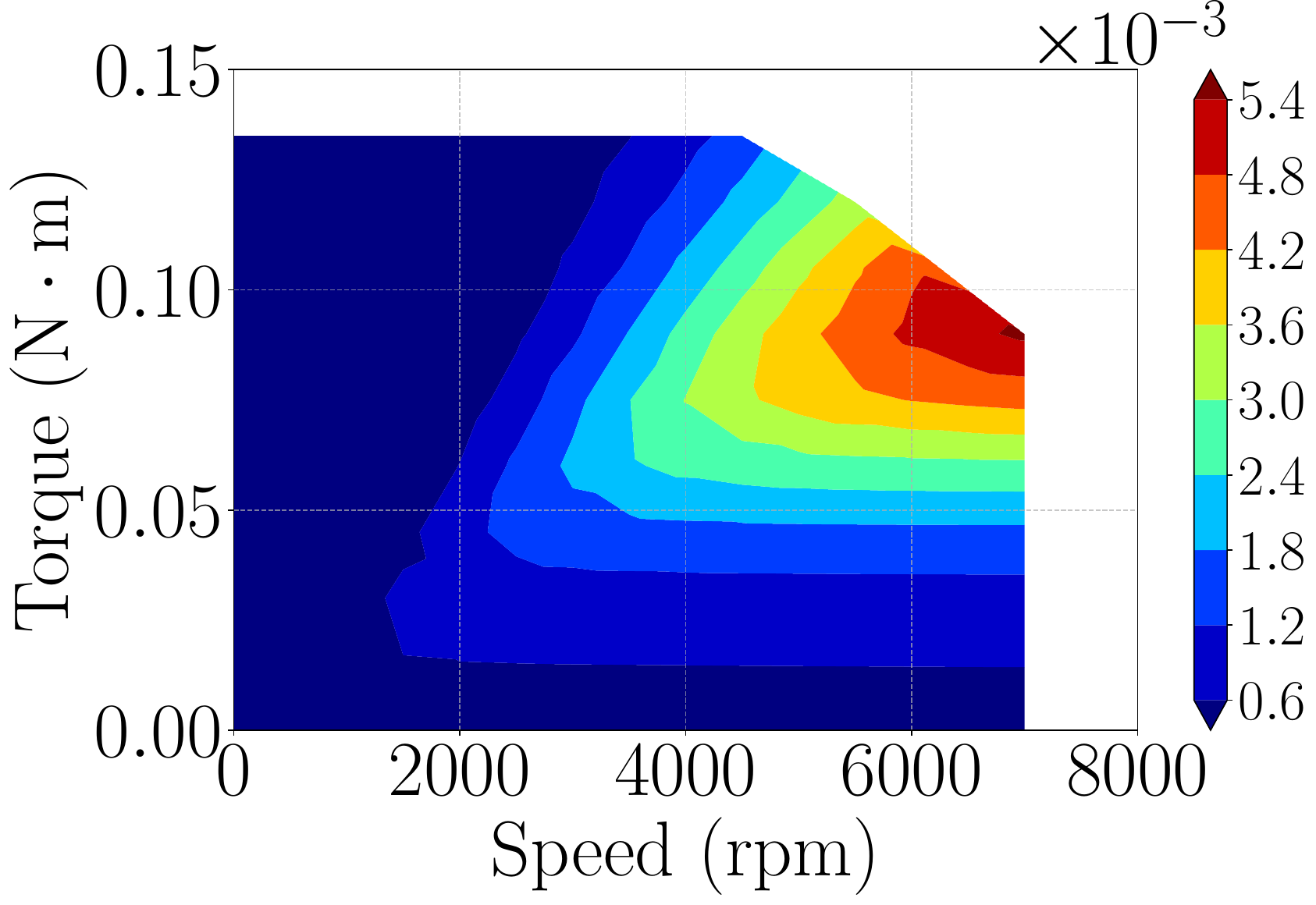}
        \caption{SF.}
    \end{subfigure}
    \\
    \begin{subfigure}[b]{0.32\textwidth}
        \centering
        \includegraphics[width=\textwidth]{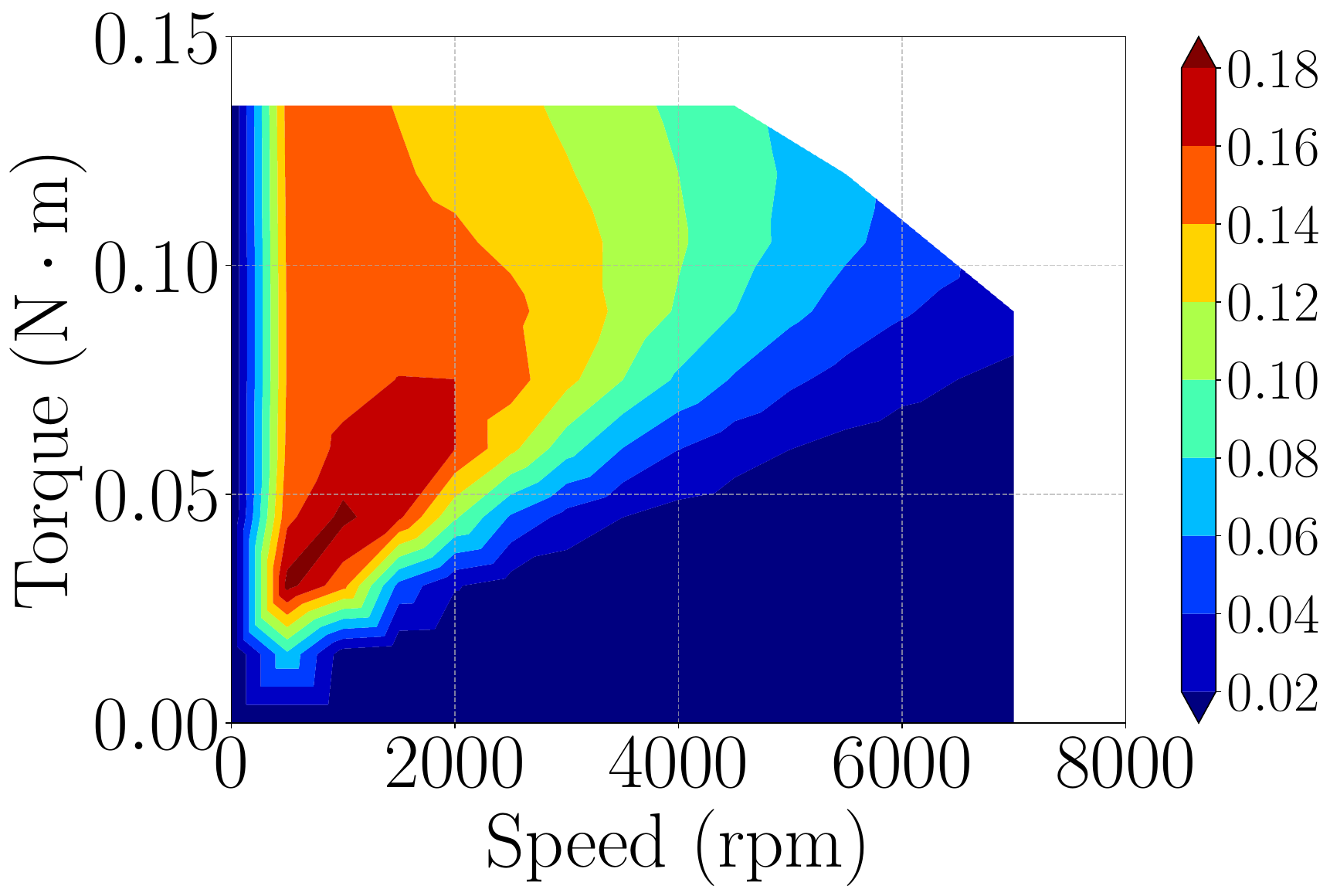}
        \caption{$R_\text{s}$.}
    \end{subfigure}
    \begin{subfigure}[b]{0.32\textwidth}
        \centering
        \includegraphics[width=\textwidth]{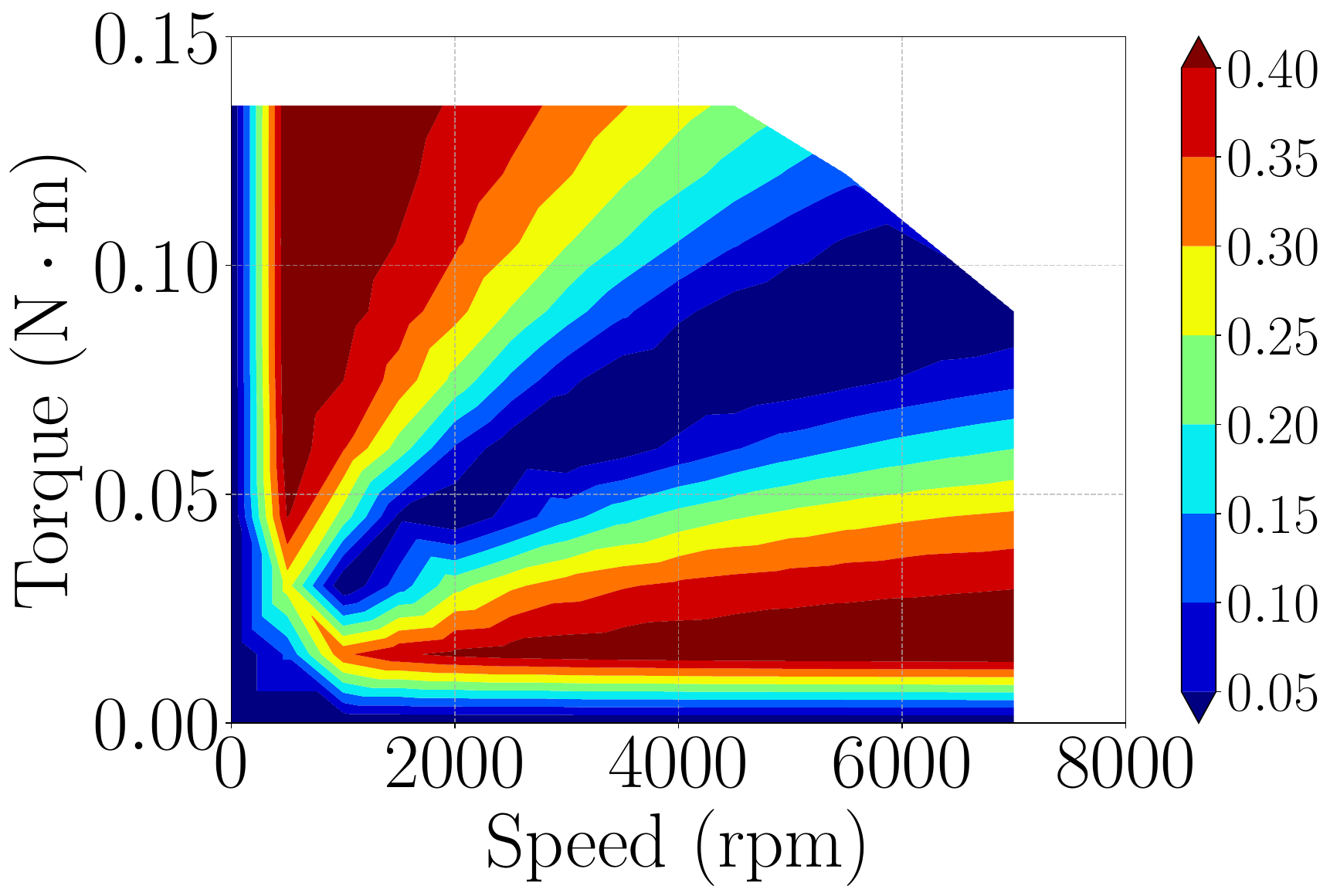}
        \caption{$B_\text{r}$.}
    \end{subfigure}
    \caption{\textcolor{black}{Elementwise Sobol' \gls{gsa} of \gls{iga} model's efficiency map (first-order indices).}}
    \label{fig:pmsm-n-grid-sa}
\end{figure*}

\begin{figure}[t!]
    \centering
    \includegraphics[width=0.75\columnwidth]{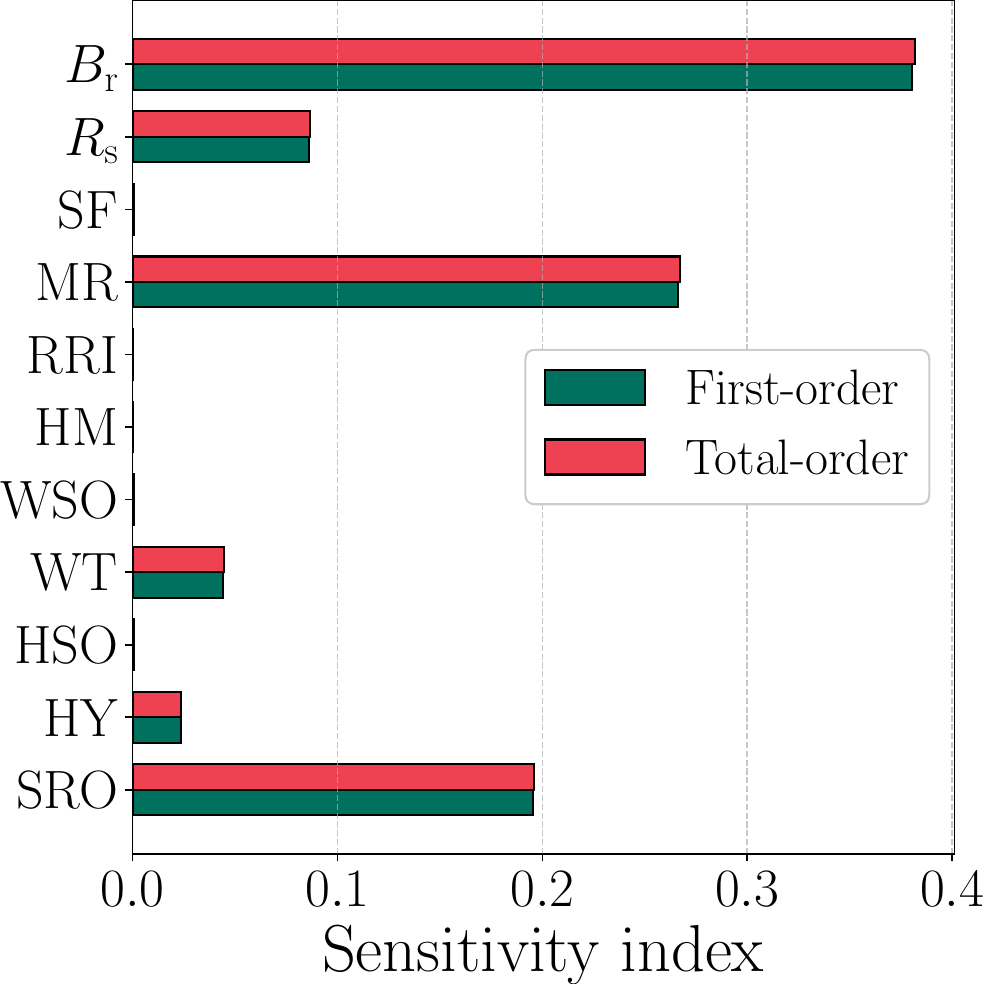}
    \caption{Multivariate \gls{gsa} of \gls{iga} model's efficiency map.}
    \label{fig:pmsm-n-grid-mvsa}
\end{figure}

Figures~\ref{fig:pmsm-n-grid-uq-mean} and \ref{fig:pmsm-n-grid-uq-std} display the mean and standard deviation of the efficiency map estimated with the adaptive \gls{pce} trained on the largest available sample.
Figure~\ref{fig:pmsm-n-error} shows the corresponding \glspl{mae} for samples on increasing size.
As can be observed, the \glspl{mae} are below $10^{-6}$ already for very small sample sizes.

Figure~\ref{fig:pmsm-n-grid-sa} presents the results of the elementwise Sobol' \gls{gsa}, while Figure~\ref{fig:pmsm-n-grid-mvsa} displays the multivariate \gls{gsa} results. 
In both cases, only first-order indices are shown; total-order indices are nearly identical, indicating negligible parameter interactions.
Both \gls{gsa} methods are in agreement that MR, SRO, HY, WT, $B_{\text{r}}$, and $R_\text{s}$ have a significant influence on efficiency variation. 
Conversely, SF, HSO, WSO, HM, and RRI exhibit negligible influence.
\textcolor{black}{
However, even though the two approaches are in agreement, multivariate \gls{gsa} presents a much clearer picture on the relative importance of the uncertain inputs parameters over the full efficiency map.
}

The results of \gls{gsa} are verified by comparing mean and standard deviation estimates based on the full \gls{iga} model with all eleven uncertain parameters, and with a reduced model where the five non-influential parameters are fixed to their nominal values. 
The low \gls{mae} values between the two models, shown in Table~\ref{tab:mad-pmsm-n}, confirm the \gls{gsa} results.

\begin{table}[t!]
\centering
\caption{\Glspl{mae} in efficiency map mean and standard deviation estimates between full and reduced \gls{iga} model.}
\label{tab:mad-pmsm-n}
\begin{tabular}{c c c}
\toprule
Fixed parameters & \gls{mae}, mean & \gls{mae}, st.d. \\
\midrule 
SF,  HSO, WSO, HM, RRI &  $5.77 \cdot 10^{-6}$ & $3.45 \cdot 10^{-6}$ \\ 
\bottomrule 
\end{tabular}
\end{table}

\section{Summary and conclusions}
\label{sec:conclusion}

This work demonstrated the use of a multivariate, variance-based \textcolor{black}{global sensitivity analysis (\gls{gsa})} method for assessing the impact of uncertain electric machine design parameters on efficiency maps and profiles.
Using two models of a \textcolor{black}{permanent magnet synchronous machine}, namely, an \textcolor{black}{equivalent circuit model} and an \textcolor{black}{isogeometric analysis} model, we showcase the shortcomings of the Sobol' \gls{gsa} method applied elementwise, as well as the benefits of multivariate \gls{gsa}. 
\textcolor{black}{
In particular, by using generalized sensitivity indices tailored to multidimensional quantities of interest (\glspl{qoi}), multivariate \gls{gsa} avoids misleading sensitivity information due to numerical artifacts, an issue often encountered in the sensitivity maps provided by the Sobol' method, and allows for a straightforward interpretation of the sensitivity results. 
}
In addition, we compare \gls{gsa} based on \textcolor{black}{Monte Carlo sampling} and \textcolor{black}{polynomial chaos expansions}, and find that the latter approach is much more computationally efficient for the selected test-cases. 
Last, we leverage the results of multivariate \gls{gsa} for model simplification, by fixing the parameters identified as non-influential to their nominal values. 
In the numerical test-cases of this work, 45\% to 50\% of the initially considered uncertain parameters can be fixed without compromising the accuracy of \textcolor{black}{uncertainty quantification} results.
We conclude that multivariate \gls{gsa} should be the preferred tool for assessing the sensitivity of efficiency maps and profiles to uncertain electric machine design parameters.

\textcolor{black}{
Naturally, the variance-based, multivariate \gls{gsa} employed in this work has its limitations.
One potential limitation is related to its underlying assumptions.
A critical assumption of variance-based \gls{gsa}, either scalar or multivariate, is that (co-)variance decomposition adequately captures all relevant uncertainty aspects. 
Another critical assumption is that of mutual independence among the uncertain input parameters. 
Should these assumptions be violated, for example, due to a heavily skewed output distribution or dependent inputs, alternative methods should be sought.}

\textcolor{black}{
Another potential limitation is related to the aggregation of sensitivity information that is inherent to multivariate \gls{gsa} methods.
Specifically, generalized sensitivity indices can potentially mask localized effects, in the sense that 
an input parameter with low impact over the full \gls{qoi} may still exert a non-negligible or even dominant influence at specific operating points. 
In essence, multivariate \gls{gsa} prioritizes interpretability and parsimony, but may possibly obscure localized sensitivities.
Conversely, the elementwise Sobol' method preserves local sensitivity information, albeit at the cost of large sensitivity data that is difficult to synthesize into practical insights.
This trade-off must be taken into account depending on the task at hand.
For the purpose of sensitivity-guided model simplification that simultaneously preserves uncertainty quantification accuracy, which was pursued in this work, multivariate \gls{gsa} is clearly the preferable choice.
However, for design scenarios where specific operating points or regions are critical, it is advisable to cross-check the elementwise sensitivity maps, or perform targeted sensitivity analysis in the particular regions of interest.
}

\textcolor{black}{
Follow-up works will investigate the applicability of multivariate \gls{gsa} to other multidimensional \glspl{qoi}, e.g., torque signals or field distributions.
For very high-dimensional \glspl{qoi}, the combination of multivariate \gls{gsa} with dimension reduction methods should be explored.
Test cases with higher-dimensional inputs could also be explored, as to establish the limits of the proposed methodology.
For example, operational and environmental uncertainties can be considered next to design uncertainties.
Last, a comparison of different multivariate \gls{gsa} methods could be useful, in order to identify their advantages and disadvantages in the context of electric machine design.}

\section*{Acknowledgments}
The authors are partially supported by the joint DFG/FWF Collaborative Research Centre CREATOR (DFG: Project-ID 492661287/TRR 361; FWF: 10.55776/F90) at TU Darmstadt, TU Graz, and JKU Linz. 
The authors thank P. K. Dhakal, K. Heidarikani, and A. Muetze (TU Graz) for providing the \gls{pmsm} \gls{ecm}, and M. Wiesheu (TU Darmstadt) for providing the \gls{iga} model.

\bibliographystyle{IEEEtran}
\bibliography{references}
\end{document}